\documentclass[useAMS,usenatbib]{mn2e}

\usepackage{epsfig}

\def\h2{H$_2$}
\def\f0{$F_0$}

\newcommand\ion[2]{#1$\;${\small\rmfamily\@Roman{#2}}\relax}%

\title[Prompt and afterglow emission of GRB~080603A]{A faint optical flash in dust-obscured GRB 080603A --
implications for GRB prompt emission mechanisms}
\author[C.~Guidorzi et~al.]{C.~Guidorzi$^{1,2}$\thanks{E-mail:guidorzi@fe.infn.it},
  S.~Kobayashi$^{2}$, D.~A.~Perley$^{3}$, G.~Vianello$^{4}$,
  J.~S.~Bloom$^{3}$, 
\newauthor P.~Chandra$^{5}$, D.~A.~Kann$^{6}$,
  W.~Li$^{3}$, C.~G.~Mundell$^{2}$, A.~Pozanenko$^{7}$,
\newauthor J.~X.~Prochaska$^{8}$, K.~Antoniuk$^{9}$, D.~Bersier$^{2}$,
  A.~V.~Filippenko$^{3}$, D.~A.~Frail$^{10}$,
\newauthor A.~Gomboc$^{11,12}$, E.~Klunko$^{13}$, A.~Melandri$^{14}$,
  S.~Mereghetti$^{4}$, A.~N.~Morgan$^{3}$, 
\newauthor P.~T.~O'Brien$^{15}$, V.~Rumyantsev$^{9}$, R.~J.~Smith$^{2}$,
  I.~A.~Steele$^{2}$, N.~R.~Tanvir$^{15}$, 
\newauthor A.~Volnova$^{16}$\\
\mbox{}\\
$^{1}$Physics Department, University of Ferrara, via Saragat 1,
  I-44122, Ferrara, Italy\\
$^{2}$Astrophysics Research Institute, Liverpool John Moores
  University, Twelve Quays House, Egerton Wharf, CH41 1LD, Birkenhead,
  UK\\
$^{3}$Department of Astronomy, University of California, Berkeley, CA
  94720-3411, USA\\
$^{4}$INAF -­ Istituto di Astrofisica Spaziale e Fisica Cosmica Milano,
  via Bassini 15, I-20133 Milano, Italy\\
$^{5}$Department of Physics, Royal Military College of Canada,
  Kingston, ON, Canada\\
$^{6}$Th\"uringer Landessternwarte Tautenburg, Sternwarte 5, D-07778
  Tautenburg, Germany\\
$^{7}$Space Research Institute (IKI), 84/32 Profsoyuznaya Str, Moscow
  117997, Russia\\
$^{8}$Department of Astronomy and Astrophysics, UCO/Lick Observatory,
  University of California, 1156 High Street, Santa Cruz, CA 95064,
  USA\\
$^{9}$SRI Crimean Astrophysical Observatory (CrAO), Nauchny, Crimea
  98409, Ukraine\\
$^{10}$National Radio Astronomy Observatory, 1003 Lopezville Road,
  Socorro, NM 87801, USA\\
$^{11}$Faculty of Mathematics and Physics, University of Ljubljana,
  Jadranska 19, SI-1000 Ljubljana, Slovenia\\
$^{12}$Centre of Excellence SPACE-SI, A\v sker\v ceva cesta 12, SI-1000
  Ljubljana, Slovenia\\
$^{13}$Institute of Solar-Terrestrial Physics, Lermontov st., 126a,
  Irkutsk 664033, Russia\\
$^{14}$INAF -- Osservatorio Astronomico di Brera, via E. Bianchi 46,
  I-23807 Merate (LC), Italy\\
$^{15}$Department of Physics and Astronomy, University of Leicester,
  University Road, Leicester LE1 7RH, UK\\
$^{16}$Sternberg Astronomical Institute, Moscow State University,
  Universitetsky pr., 13, Moscow 119992, Russia
}

\begin{document}

\date{\today}


\maketitle

\label{firstpage}

\begin{abstract}

We report the detection of a faint optical flash by the 2-m Faulkes
Telescope North simultaneously with the second of two prompt
$\gamma$-ray pulses in INTEGRAL gamma-ray burst (GRB) 080603A,
beginning at $t_{\rm rest} = 37$~s after the onset of the GRB.  This
optical flash appears to be distinct from the subsequent emerging
afterglow emission, for which we present comprehensive broadband radio
to X-ray light curves to 13 days post-burst and rigorously test the
standard fireball model.  The intrinsic extinction toward GRB~080603A
is high ($A_{V,z} = 0.8$~mag), and the well-sampled
X-ray-to-near-infrared spectral energy distribution is interesting in
requiring an LMC2 extinction profile, in contrast to the majority of
GRBs.  Comparison of the $\gamma$-ray and extinction-corrected optical
flux densities of the flash rules out an inverse-Compton origin for
the prompt $\gamma$-rays; instead, we suggest that the optical flash
could originate from the inhomogeneity of the relativistic flow.  In
this scenario, a large velocity irregularity in the flow produces the
prompt $\gamma$-rays, followed by a milder internal shock at a larger
radius that would cause the optical flash.  Flat $\gamma$-ray spectra,
roughly $F \propto \nu^{-0.1}$, are observed in many GRBs.  If the
flat spectrum extends down to the optical band in GRB~080603A, the
optical flare could be explained as the low-energy tail of the
$\gamma$-ray emission.  If this is indeed the case, it provides an
important clue to understanding the nature of the emission process in
the prompt phase of GRBs and highlights the importance of deep ($R >
20$~mag), rapid follow-up observations capable of detecting faint,
prompt optical emission.

\end{abstract}


\section{Introduction}
\label{sec:intro}

The exact mechanism that produces the prompt radiation of a gamma-ray
burst (GRB) is still unknown. As a nonthermal process, synchrotron and
inverse Compton (IC) are the main candidates. The former can
successfully account for most of the afterglow emission evolution, and
is naturally expected from shock-accelerated electrons. As such, this
has also been considered for explaining the $\gamma$-ray prompt
emission itself. However, the energy spectrum of a GRB, usually
modelled with a Band function \citep{Band93}, is such that the typical
value for the low-energy photon index ($\alpha$) violates the
so-called ``synchrotron death line'' ($\alpha = -2/3$) for a sizable
fraction of cases \citep{Preece98,Guiriec10,Guidorzi11}.  In addition,
the value generally observed, $\alpha \approx -1$, differs from the
value of $-3/2$ expected for rapidly cooling electrons (the so-called
``fast-cooling death line''; \citealt{Ghisellini00}). Although under
some assumptions most GRB spectra could be reconciled with a
synchrotron origin (e.g., \citealt{Lloyd00,Daigne11}), the question of
whether it is the dominant process in the GRB production remains
unanswered.

On the other hand, IC has been considered as a possible alternative,
such as synchrotron self-Compton (SSC; \citealt{Kumar08}), especially
when the prompt optical emission is very bright \citep{Racusin08}.  
IC as the source of $\gamma$-rays requires a soft component in the
infrared (IR) through ultraviolet (UV) range for providing the seed
photons; this in turn means that the second IC component peaks in the
GeV--TeV range, potentially implying an ``energy crisis'' problem
\citep{Piran09}.  Combining prompt optical and $\gamma$-ray
measurements, together with the wealth of information derived from the
broadband modelling of the early-to-late time afterglow, offers a
direct way test for IC as the mechanism for the GRB production.

The standard afterglow model (see, e.g., \citealt{Meszaros06} for a
review) is rather successful in modelling the temporal and spectral
evolution of GRB afterglows. However, models often require
modifications, such as energy injections from long-lived internal
engines or density enhancements in the surrounding medium, and for a
sizable fraction of cases, even these options cannot provide a fully
satisfactory explanation (e.g., \citealt{Melandri08}).

Long-duration GRBs are also probes of the interstellar and
intergalactic medium, and of cosmic star formation history up to
redshift $z \approx 8$ \citep{Salvaterra09,Tanvir09}, potentially
exploring the reionization epoch \citep{Kistler09,Robertson10}.
Spectral energy distribution (SED) studies and spectroscopic
observations can help to shed light, for instance, on the redshift
evolution of dust, gas content, and metallicity of the host galaxies
of GRBs as well as of the local region within the host (e.g.,
\citealt{Prochaska07}).  In addition, they help to identify the
crucial physical parameters which favour the production of GRBs.
Dust-extinction modelling for a sample of GRB afterglows with
well-sampled SEDs has shown that most cases can be described with
Small Magellanic Cloud (SMC) profiles \citep{Kann10} having little
evidence for the 2175\,\AA\ bump seen in the Milky Way extinction
curve, except for very few cases
\citep{Kruhler08,Prochaska09,Eliasdottir09,Perley11}.

No direct link has been found between the properties of the prompt
emission and those of the circumburst environment surrounding the GRBs
and of the host galaxy, such as metallicity \citep{Levesque10}.
Nevertheless, a detailed picture of the properties that can be derived
from the broadband afterglow modelling --- the dust content and
features along the sightline to the GRB within the host galaxy --- are
crucial to provide a self-consistent description of the entire GRB
phenomenon, and for unveiling the yet unknown connections between the
GRB itself and its birthplace and (to some extent) progenitor.

This paper reports comprehensive analysis and discussion of the
multi-wavelength dataset collected on the long-duration GRB~080603A
detected by INTEGRAL \citep{Winkler03} in light of the current
standard fireball model.  This GRB provides an ideal test bed because
it had an optical flash simultaneous with the prompt emission, and we
recorded the broadband afterglow SED and its evolution.  Our dataset
includes INTEGRAL data of the $\gamma$-ray prompt emission itself, as
well as multi-filter photometric and spectroscopic data of the
near-infrared/optical afterglow and of the host galaxy.  In addition,
we analysed the X-ray afterglow data, discovered in the 0.3--10~keV
band with {\it Swift}/XRT \citep{Sbarufatti08b}, from 3~hours to
7~days after the burst. We also include data from the Very Large Array
(VLA), taken from 2 to 13 days post burst, in which the radio
afterglow was detected.

Throughout the paper, times are UT and are given relative to the GRB
onset time as observed with INTEGRAL, which corresponds to June 3,
2008, 11:18:11~UT. The convention $F(\nu,t) \propto
\nu^{-\beta}\,t^{-\alpha}$ is followed, where the energy index $\beta$
is related to the photon index by $\Gamma=\beta + 1$.  We adopted the
standard cosmological model: $H_0=71$\,km\,s$^{-1}$\,Mpc$^{-1}$,
$\Omega_\Lambda=0.73$, $\Omega_{\rm M}=0.27$ \citep{Spergel03}.

All of the quoted errors are given at 90\% confidence level for one
interesting parameter ($\Delta \chi^2 = 2.706$), unless stated
otherwise.

\section{Observations}
\label{sec:obs}

GRB~080603A was detected with the INTEGRAL/IBIS instrument and
localised in real time by the INTEGRAL Burst Alert System (IBAS;
\citealt{Mereghetti03}) with an accuracy of $3.2\arcmin$.  The
$\gamma$-ray prompt emission in the 20--200~keV energy band lasted
about 150~s.  A quick-look analysis gave a peak flux of
0.5~ph~cm$^{-2}$~s$^{-1}$, a fluence of about $10^{-6}$~erg~cm$^{-2}$,
and burst coordinates $\alpha$(J2000) $= 18^{\rm h} 37^{\rm m}
38\fs2$, $\delta$(J2000) $= +62^\circ 44\arcmin 06\arcsec$ with an
error radius of $2\arcmin$ \citep{Paizis08}.

The Faulkes Telescope North (FTN) promptly reacted to the IBAS alert
and began observing at 11:19:51, which was 100~s after the burst onset
time and while prompt $\gamma$-ray emission was still ongoing.  The
automatic GRB pipeline did not identify any candidate; consequently, a
pre-programmed $BVRi'$ observation sequence with increasingly longer
exposure durations was carried out \citep{Guidorzi06}. However, visual
inspection of the frames revealed the presence of an uncatalogued,
variable object at $\alpha$(J2000) $=18^{\rm h} 37^{\rm m} 38\fs1$,
$\delta$(J2000) $=+62^\circ 44\arcmin 39\farcs4$ with $R = 19.6$~mag
at $t = 7.37$~min, calibrated against nearby USNOB--1.0 stars
\citep{Gomboc08}.  FTN observations continued for 3~hours.

The GRB alert system of the Katzman Automatic Imaging Telescope (KAIT;
\citealt{Li03}) at Lick Observatory also promptly reacted to the
INTEGRAL alert and independently detected the optical counterpart at a
position consistent with that of the FTN, reporting $I = 18.7$~mag at
10.7~min in unfiltered and $I$-band images \citep{Chornock08}.  Other
robotic telescopes also reported the discovery of the afterglow
\citep{Milne08}.

We began spectroscopic observations of the optical afterglow with the
GMOS dual spectrometer at the Gemini-North 8-m telescope starting at
time 13:24, identifying several absorption features at a common
redshift of $z = 1.68742$, in agreement with preliminary reports
\citep{Perley08a}.

We kept monitoring the evolution of the optical afterglow with the
1.34~m Schmidt telescope of the Th\"uringer Landessternwarte (TLS),
the Liverpool Telescope (LT), the Keck-I 10-m telescope, the
Zeiss--1000 and AZT--11 telescopes at the Crimean Astrophysical
Observatory (CrAO), and the AZT--33IK telescope at Sayan Observatory
up to 4 days post burst
\citep{Kann08,Rumyantsev08a,Rumyantsev08b,Perley08b,Klunko08,Rumyantsev08}.
From 19 to 22~hours post burst we observed the afterglow at
near-infrared (NIR) wavelengths with the Peters Automated Infrared
Imaging Telescope (PAIRITEL) through $JHK$ filters \citep{Miller08}.
Our last detection of the optical afterglow was obtained with the
Low Resolution Imaging Spectrometer (LRIS; \citealt{Oke95})
4 days post burst.
Two months after the burst we observed the same field with
Keck/LRIS using $u'g'RI$ filters and detected the host galaxy; these
observations were subsequently used to subtract the host contribution
when the afterglow became comparably faint.

{\it Swift}/XRT began observing GRB~080603A from 2.9 to 5.9~hours and
reobserved it from 2.5 to 7.0~days after the burst. The X-ray
afterglow was clearly identified at position $\alpha$(J2000) $=18^{\rm
  h} 37^{\rm m} 38\fs06$, $\delta$(J2000) $=+62^\circ 44\arcmin
40\farcs1$ with an error radius of $1\farcs9$ \citep{Sbarufatti08a}.
During the same time interval {\it Swift}/UVOT detected the fading
afterglow in the $V$ band \citep{Kuin08,Sbarufatti08b}.

Finally, we discovered the radio counterpart with the VLA at 4.86 and
8.46~GHz, initially at the latter frequency with a possible detection
at 1.9~days at a flux density of $116 \pm 41$~$\mu$Jy
\citep{Chandra08a}. The detection was confirmed at 3.95~days
\citep{Chandra08b}. Observations continued as late as 13~days post
burst.

The Galactic reddening along the direction to the GRB is $E_{B-V} =
0.044$~mag \citep{Schlegel98}. The Galactic extinction in each filter
has been estimated through the NASA/IPAC Extragalactic Database
extinction
calculator\footnote{http://nedwww.ipac.caltech.edu/forms/calculator.html.}.
Specifically, the extinction in each filter is derived through the
parametrisation by \citet{Cardelli89}: $A_{u'}=0.23$, $A_B=0.19$,
$A_{g'}=0.18$, $A_V=0.14$, $A_{r'}=A_R=0.12$, $A_{i'}=0.09$,
$A_I=0.08$, $A_J=0.04$, $A_H=0.025$, and $A_K=0.02$~mag.

\section{Data reduction and analysis}
\label{sec:an}

\subsection{Gamma-ray data}
\label{sec:gamma}

\begin{figure}
\centering
\includegraphics[width=8.5cm]{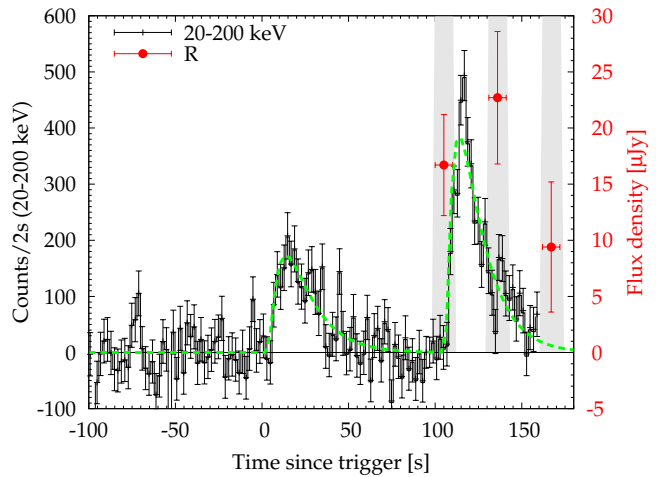}
\caption{INTEGRAL light curve in the 20--200~keV energy band
  (left-hand ordinate axis). The shaded areas display the time
  intervals of the first optical measurements with the FTN; the
  corresponding $R$-band flux densities are shown with filled circles
  (right-hand ordinate axis).  The dashed line shows the best-fitting
  models of the $\gamma$-ray pulses as obtained with the model by
  \citet{Norris05}.}
\label{f:lc_gamma}
\end{figure}

Figure~\ref{f:lc_gamma} shows the 20--200~keV background-subtracted
time profile of GRB~080603A recorded by the IBIS/ISGRI detector
\citep{Lebrun03}.  The profile consists of two very similar pulses of
duration 30-s, peaking at 14~s and 114~s. A combination of two
fast-rise-exponential-decline (FRED)-like pulses as modelled by
\citet{Norris05} gives a satisfactory result ($\chi^2/{\rm  dof}=94.4/81$),
as shown by the dashed line in Figure~\ref{f:lc_gamma}.
The parameters used are the peak time $t_{\rm peak}$, the peak intensity $A$,
the rise and decay times $\tau_{\rm r}$ and $\tau_{\rm d}$, the pulse
width $w$, and the asymmetry $k$.
Their best-fitting values are reported in Table~\ref{tab:N05}.
Apart from the peak of the second pulse, which is roughly twice as
intense as that of the first, the two pulses share very similar
temporal properties: rise and decay times around 7 and 20~s,
respectively, with a corresponding decay-to-rise ratio around a factor
of 3, very typical of classical FREDs \citep{Norris96}.
%
%

Two time-integrated spectra, one for each pulse, show no evidence for
spectral evolution: a simple power law can fit both spectra with a
$\gamma$-ray photon index $\Gamma_\gamma \approx 1.6$.
Table~\ref{tab:gamma} reports the best-fitting spectral parameters.
The 20--200~keV total fluence is $(1.1 \pm 0.2) \times
10^{-6}$~erg~cm$^{-2}$, in agreement with preliminary reports
\citep{Paizis08}.  The value of $\Gamma_\gamma = 1.6$ lying between
the typical low-energy and high-energy photon indices of GRB prompt
emission spectra (e.g., \citealt{Kaneko06,Sakamoto09}) suggests that
the peak energy, $E_{\rm p}$, is likely to lie within the 20--200~keV
energy band.  In the context of the sample of GRBs detected with ISGRI
and BAT \citep{Vianello09}, the fluence of GRB~080603A makes it a
typical burst.

The corresponding flux-density curve shown in
Figure~\ref{f:lc_panchro} was found to refer to 84~keV; this is the
energy at which the energy spectrum with $\beta_\gamma = \Gamma_\gamma
- 1 = 0.6$ has the same value as that averaged over the 20--200~keV
range.

\begin{table*}
\centering
\caption{Best-fitting parameters of the time profile of the
$\gamma$-ray pulses as seen in the 20--200~keV band.}
\label{tab:N05}
\begin{tabular}{lcccccc}
\hline
 Pulse  & $t_{\rm peak}$   & $A$(at 84 keV) & $\tau_{\rm r}$  & $\tau_{\rm d}$  & $w$   & $k$\\
        &    (s)       & ($\mu$Jy)&    (s)         &      (s)      &  (s)  &\\
\hline
    1   &  $13.7\pm1.9$ & $18\pm2$ & $8.0\pm2.0$ & $21.8\pm3.5$ & $29.8\pm3.8$ & $0.46\pm0.12$\\
    2   &  $113.8\pm0.9$ & $41\pm2$ & $6.2\pm1.1$ & $19.4\pm1.6$ & $25.7\pm1.8$ & $0.51\pm0.08$\\
\hline
\end{tabular}
\end{table*}

\begin{table*}
\centering
\caption{Best-fitting parameters of the energy spectra of the 
$\gamma$-ray prompt emission in the 20--200~keV band. The model is 
a power-law and $\Gamma_\gamma$ is the photon index.}
\label{tab:gamma}
\begin{tabular}{cccccc}
\hline
 Pulse  & Time interval   & $\Gamma_\gamma$   & Average flux            & $\chi^2/{\rm dof}$  & Fluence\\
        &      (s)        &    & ($10^{-8}$~erg~cm$^{-2}$~s$^{-1}$)&     & ($10^{-7}$~erg~cm$^{-2}$)\\
\hline
   1    &   3--38       & $1.6\pm0.2$           & $1.4_{-0.6}^{+0.14}$   &  $60/64$  & $4.1\pm1.3$\\
   2    &   100--140    & $1.65_{-0.16}^{+0.18}$   & $1.9_{-0.6}^{+0.16}$   &  $39/30$  & $6.7\pm1.5$\\
\hline
\end{tabular}
\end{table*}

Despite the unknown value of $E_{\rm p}$, we can provide a
conservative estimate of the isotropic-equivalent radiated energy
$E_{\rm iso}$ in the GRB rest-frame 1--$10^4$~keV energy band: we
assume that $E_{\rm p}$ lies either within or close to the 20--200~keV
energy range. In the former case, we use the logarithmic average,
$E_{\rm p} = 60$~keV, while in the latter case we consider the values
10~keV and 400~keV as the lower and upper boundary, respectively.
These values correspond to a 0.3 logarithmic shift from the
corresponding boundary, the logarithmic bandwidth being 1. In
calculating the fluence in the rest-frame 1--$10^4$~keV band, the
K-correction factor is $2.8 \pm 0.8$, where the uncertainty accounts
for the different $E_{\rm p}$ assumed and where we adopted the typical
Band function with $\alpha_{\rm B} = -1$ and $\beta_{\rm B} = -2.3$
\citep{Kaneko06}.  As a result, we estimate $E_{\rm iso} = (2.2 \pm
0.8) \times 10^{52}$~erg and the intrinsic peak energy $E_{\rm p,i} =
160_{-130}^{+920}$~keV; these values are broadly consistent with the
$E_{\rm p,i}$--$E_{\rm iso}$ relation \citep{Amati02,Amati10},
although the poor accuracy on $E_{\rm p,i}$ is not very constraining.

\subsection{X-ray data}
\label{sec:xray}

The {\it Swift}/XRT began observing GRB~080603A on 2008 June 03 at
14:11:19, about 10.4~ks after the burst, and ended on 2008 June 10 at
11:44:56, with a total net exposure of 17.8~ks in photon counting (PC)
mode spread over 6.9 days.  The XRT data were processed using the
FTOOLS software package (v. 6.7) distributed within HEASOFT. We ran
the task {\sc xrtpipeline} (v.0.12.1), applying calibration and
standard filtering and screening criteria. Data were acquired only in
PC mode due to the faintness of the source. Events with grades 0--12
were selected. The XRT analysis was performed in the 0.3--10~keV
energy band.

Source photons were extracted from a circular region centred on the
final XRT position \citep{Sbarufatti08b} and with a radius of
20~pixels (1~pixel $= 2\farcs36$), and were point-spread function
(PSF) renormalised.  Background photons were extracted from nearby
circular regions with a total area of $22.7 \times 10^3$~pixels away
from any source present in the field.  No pile-up correction was
required because of the low count rate ($\la 0.1$~count~s$^{-1}$) of
the source from the beginning of the XRT observations.  When the count
rate dropped below $\sim 10^{-2}$~count~s$^{-1}$, we made use of 
{\sc ximage} with the tool {\sc sosta}, which corrects for vignetting,
exposure variations, and PSF losses within an optimised box, using the
same background region.

We extracted the 0.3--10~keV energy spectrum in the time interval from
10.4 to 21.1~ks; later observations did not allow us to collect enough
photons to ensure the extraction of another meaningful spectrum.
Source and background spectra were extracted from the same regions as
those used for the light curve.  Spectral channels were grouped so as
to have at least 20 counts per bin. The ancillary response files were
generated using the task {\sc xrtmkarf}. Spectral fitting was
performed with {\sc xspec} (v. 12.5).  The spectrum can be modelled
with an absorbed power law with the combination of {\sc xspec} models
{\sc wabs zwabs pow}, based on the photoelectric cross section by
\citet{Morrison83}.  Results of the best-fit parameters are reported
in Table~\ref{tab:sed}. The Galactic neutral hydrogen column density
along the GRB direction was fixed to the value determined from 21~cm
line radio surveys: $N_{\rm HI}^{\rm Gal} = 4.7 \times
10^{20}$~cm$^{-2}$ \citep{Kalberla05}.  The additional X-ray
absorption, modelled in the GRB rest frame, was found to be
$N_{{\rm HI},z} = 6.6_{-4.6}^{+6.2} \times 10^{21}$~cm$^{-2}$, very typical
of X-ray afterglow spectra (e.g., \citealt{Campana10}).  The X-ray
photon index in the 0.3--10~keV energy band is $\Gamma_X = 2.3 \pm
0.3$.

The X-ray unabsorbed flux light curve was derived from the rate curve
by assuming the same counts-to-energy factor ($5.4 \times 10^{-11}$
erg~cm$^{-2}$~s$^{-1}$~count$^{-1}$) obtained from the spectrum
described above.  This implicitly relies on the lack of strong
spectral evolution from $\sim 10$~ks onward; although such an
assumption cannot be proven due to the paucity of photons at late
times, this is in agreement with what is observed for most GRBs (e.g.,
\citealt{Evans09}).  Finally, the flux-density curve shown in
Figure~\ref{f:lc_panchro} was calculated at 2.4~keV, the energy at
which the energy spectrum with $\beta_X = \Gamma_X - 1 = 1.3$ has the
same value as that averaged over the 0.3--10~keV range.

\subsection{Infrared/optical data}
\label{sec:opt}

The FTN carried out robotically triggered observations between 100~s
and 190~min.  During the detection mode, consisting of the first $3
\times 10$~s frames in the $R$ band, the optical afterglow was too
faint to be automatically identified, so the GRB pipeline LT--TRAP
\citep{Guidorzi06} triggered the multi-filter ($BVRi'$) observation
sequence with increasingly longer exposures.  However, a quick visual
inspection of the data led to the identification of an uncatalogued
and variable source proposed to be the afterglow candidate
\citep{Gomboc08}.  Our best estimate for the optical afterglow
position is $\alpha$(J2000) $=18^{\rm h} 37^{\rm m} 38\fs05$,
$\delta$(J2000) $=+62^\circ 44\arcmin 39\farcs4$ with an error radius
of $0\farcs5$, and it lies within the final XRT error circle.

The afterglow observations with the FTN occurred during the onset of
the second $\gamma$-ray pulse; concurrently, a faint optical flash
with an $R$ magnitude varying within the 20--21 range was observed
(Fig.~\ref{f:lc_gamma}). The flash was soon followed by a steep rise
and a broad plateau around $10^3$~s, at the end of which a smooth
transition to a typical power-law decay with index around 1 took
place (Fig.~\ref{f:lc_panchro}).

Later observations were carried out with the LT from 15.5 to
17.1~hours with the Sloan Digital Sky Survey (SDSS) $r'$ and $i'$
filters, as well as with the FTN from 21.4 to 26.4~hours in $i'$.

Calibration of the $BVRr'i'$ frames was performed by comparing with
the magnitudes of four nonsaturated field stars. The corresponding
zero points were determined through the observations of
\citet{Landolt92} field stars for which \citet{Smith02} provide an
SDSS calibration.  In both cases the zero points were stable during
the night, showing fluctuations as large as 0.02--0.03~mag in the
worst cases.  Finally, we corrected for the airmass.  Both aperture
and PSF photometry was systematically carried out using the Starlink
{\sc gaia}
software\footnote{http://starlink.jach.hawaii.edu/starlink.}, making
sure that both gave consistent results within the uncertainties.
Magnitudes were converted into flux densities ($\mu$Jy) following
\citet{Fukugita95,Fukugita96}. Results are reported in
Table~\ref{tab:phot}; magnitudes are corrected for airmass, while flux
densities are also corrected for Galactic reddening.

%
\begin{figure*}
\centering
\includegraphics[width=4.5cm]{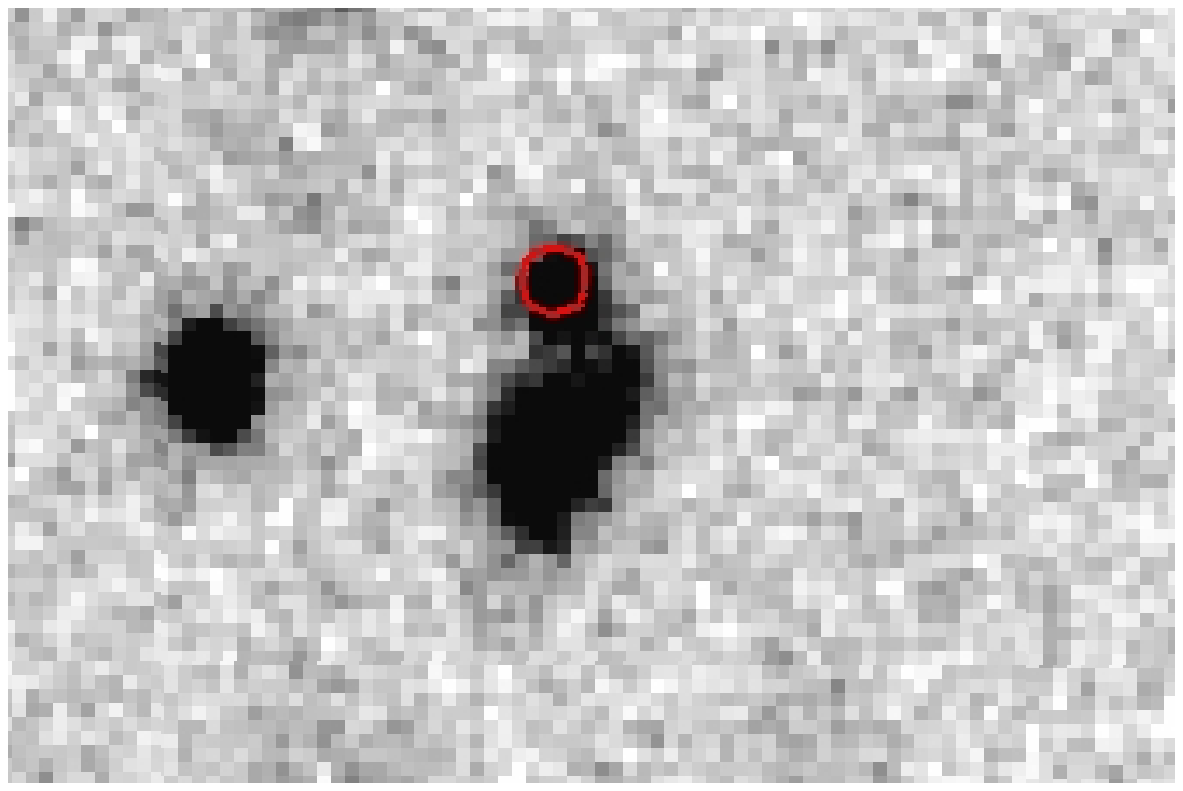}
\includegraphics[width=4.5cm]{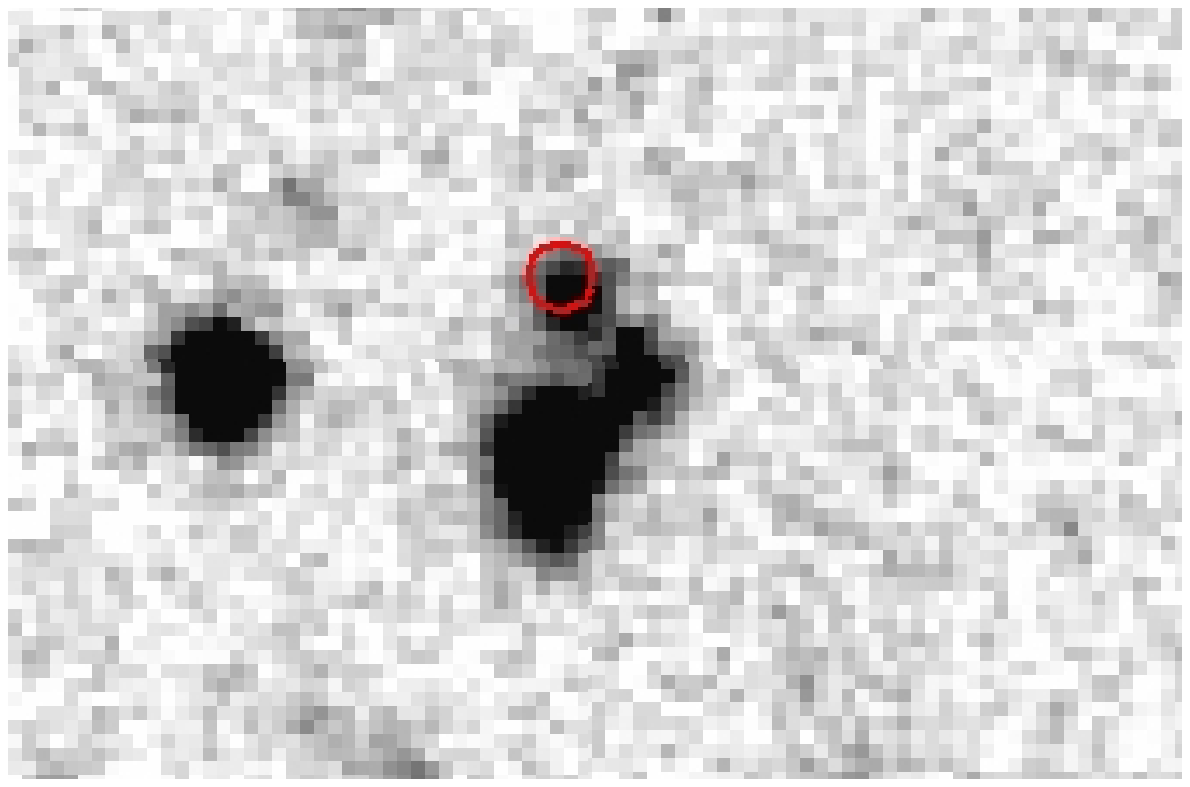}
\includegraphics[width=4.5cm]{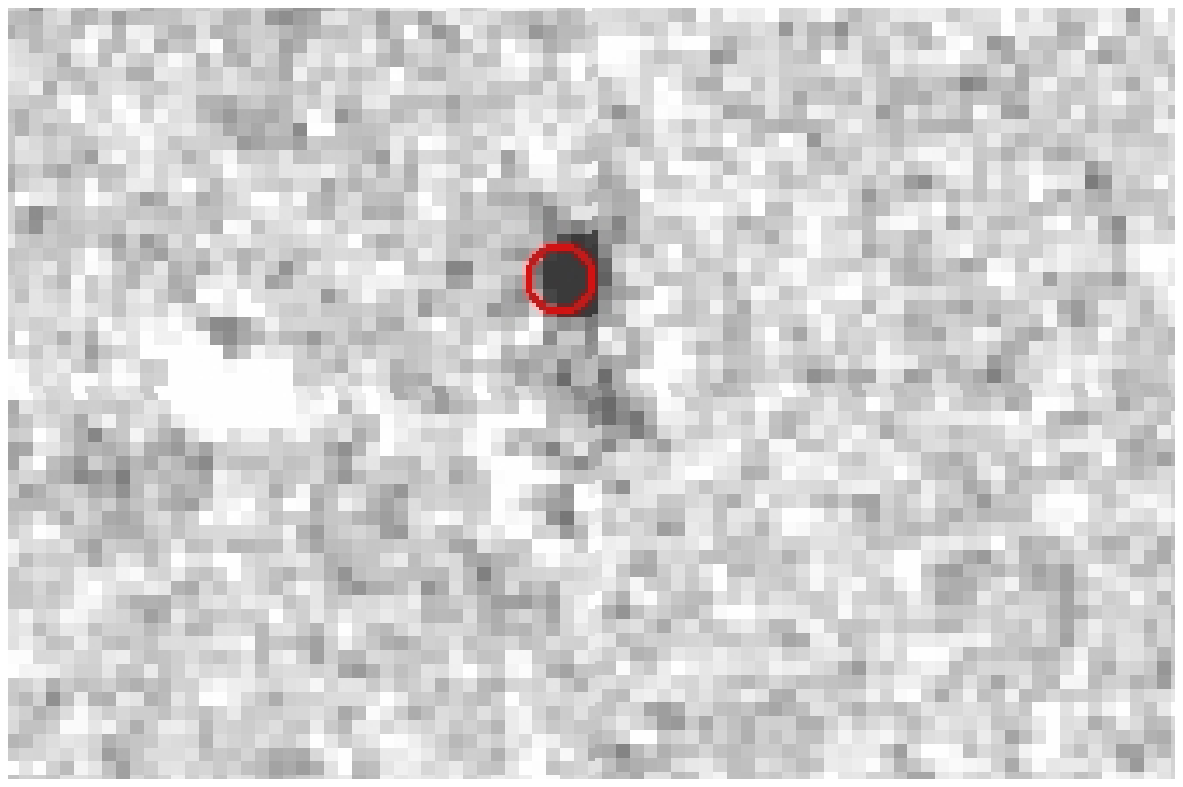}
\caption{Keck-LRIS $R$-band images of the crowded field of
  GRB~080603A.  The circle is our best position of the optical
  afterglow obtained from early FTN frames with $0\farcs5$ error
  radius.  {\em Left panel}: taken at 4.1~days; the afterglow is still
  detected.  {\em Middle panel}: taken two months after the burst; the
  host galaxy is clearly visible within the error circle, while the
  afterglow is no longer detectable. {\em Right panel}: subtracted
  image.}
\label{f:keck}
\end{figure*}
%

KAIT observations began at $t = 253$~s in the $V$, $I$, and unfiltered
bands; the pre-programmed exposure durations generally increased with
time. The earliest firm unfiltered detection of the optical afterglow
occurred at $t \approx 11$~min \citep{Chornock08}, independent of the
FTN detection.  Successive stacked frames also gave a later detection
with the $I$ filter, while the $V$ filter only provided either upper
limits or very marginal detections, as reported in
Table~\ref{tab:phot}.  For the $V$ filter, we also considered the UVOT
photometric points provided by \citet{Sbarufatti08b}.  We used some of
the four FTN field stars to calibrate the KAIT field; for the
unfiltered frames we adopted the zero point of the $R$ band.  The last
useful frame obtained with KAIT was acquired at 20~min.  Despite the
large uncertainties, both the measured values and upper limits are in
agreement with the contemporaneous values obtained with the FTN, as
shown by Figure~\ref{f:lc_panchro}.

At a midpoint time of 1.55~days we observed GRB~080603A with the
1.34-m Schmidt TLS, obtaining a total of $6 \times 600$~s images in
the $R$ band \citep{Kann08}. Stacking all six frames, the afterglow is
detected.  In order to subtract the contribution from the extended
object later identified as the host galaxy and properly account for
the crowded field, another frame of the same field was taken on 2008
August 26 and used for image subtraction with the {\sc ISIS}
package\footnote{http://www2.iap.fr/users/alard/package.html.}
\citep{Alard98}.  The brightness, estimated with both aperture
photometry and SExtractor (v.~2.5.0; \citealt{Bertin96}), is $R = 22.1
\pm 0.3$ mag.

Other late-time observations with $R$ filters were obtained with the
1.25-m AZT--11 (at 0.36~days) and 1-m Zeiss--1000 (Z1000, at 1.6 and
3.6 days) telescopes of the Crimean Astrophysical Observatory (CRAO;
\citealt{Rumyantsev08a,Rumyantsev08b,Rumyantsev08}), and with the
1.5-m AZT--33IK telescope (at 0.30~days) of the Sayan Observatory
\citep{Klunko08}.  In all cases the afterglow was detected, except for
the last observation at 3.6~days, which provided an upper limit of $R
> 22.9$~mag.  For the same reasons as in the case of the TLS frame, we
had to correct the measured $R$ magnitude at 1.6~days for the
host-galaxy contribution.  Given the lack of late-time images, we
merely subtracted the host-galaxy flux contribution as estimated with
Keck (see below); this turned into a shift of $0.2$~mag in $R$,
comparable with the uncertainty affecting the measurement itself,
which in the end was $R = 22.52 \pm 0.25$~mag.

The 1.3-m PAIRITEL started observing the afterglow of GRB~080603A at
18.9~hours with $JHK_s$ filters. Photometric calibration was done
against seven nearby 2MASS stars; magnitudes were estimated with both
aperture and PSF photometry under {\sc gaia}.  The NIR afterglow
counterpart is clearly detected in all filters in two mosaic frames
centred at 0.82 and 0.95~days with 2822~s and 4363~s total exposures,
respectively. Our estimates agree within uncertainties with the
preliminary results \citep{Miller08}.

\subsubsection{Late-time host galaxy observations}
\label{sec:opt_host}

We used the Keck LRIS to observe GRB~080603A at two different epochs.
The first run was taken on 2008 June 7, between times 12:31 and 12:47
with $R$ (total exposure 690~s) and $g'$ (total exposure 785~s)
filters. The average airmass was 1.38 and the D560 dichroic was used.
The afterglow was clearly detected with both filters at the FTN
position.

The same field was reobserved on 2008 August 2, with $R$ (total
exposure 930~s) and $g'$ (total exposure 1110~s) filters.  The average
airmass was 1.58 and the D560 dichroic was used. The next night, we
observed with the $u'$ and $I$ filters (total exposure 720~s each),
using the D680 dichroic, with an average airmass of 2.00.

Calibration was based on the identification of seven common field
stars having $R = 20$--23~mag, bright enough to be accurately measured
in the stacked images of the FTN+LT and faint enough to avoid
saturation in the Keck images.  Transformations between the SDSS and
Johnson-Cousins systems for this set of faint stars were done
following \citet{Jordi06}. The scatter in the zero point of each
filter was incorporated into the uncertainties.

In order to subtract the contribution of the host galaxy from the
afterglow flux observed at 4.1~days, we performed image subtraction
with {\sc isis}.  Figure~\ref{f:keck} shows the result in the $R$
band: the afterglow plus host at 4.1~days, the host two months later,
and the difference between the two are shown in the left, middle, and
right panels, respectively.  From image subtraction, we estimated $R =
24.17 \pm 0.14$ and $R_{\rm host} = 24.03 \pm 0.12$~mag for the
afterglow and the host, respectively. Correspondingly, we found $g' =
24.95 \pm 0.20$ and $g'_{\rm host} = 24.78 \pm 0.16$~mag.  We also
used the late-time $R$ and $I$ frames to subtract the host
contribution from the observations performed with the FTN and LT at $t
\approx 1$~day, when the afterglow brightness was around 21 mag in the
same filter. In the two cases this turned into a shift of $\sim
0.1$~mag, similar to the corresponding statistical uncertainties. For
the same reasons, we did not correct the comparably faint afterglow
magnitudes of the optical flash seen in the first FTN images, because
of their relatively large uncertainties.

\begin{table}
\centering
\caption{Host-galaxy magnitudes.}
\label{tab:host_mag}
\begin{tabular}{lrll}
\hline
Filter & $\lambda_{\rm eff}$ & Magnitude & Corrected$^{\mathrm{a}}$\\
       & (\AA)             &           &                       \\
\hline
$u'$ & 3450 & $25.6\pm0.4$   & $25.4\pm0.4$\\
$g'$ & 4731 & $24.78\pm0.16$ & $24.60\pm0.16$\\
$R$  & 6417 & $24.03\pm0.12$ & $23.91\pm0.12$\\
$I$  & 7599 & $23.98\pm0.15^{\mathrm{b}}$ & $23.89\pm0.15^{\mathrm{b}}$\\
\hline
\end{tabular}
\begin{list}{}{}
\item[$^{\mathrm{a}}$]For Galactic extinction.
\item[$^{\mathrm{b}}$]Calibration was done against $i'$ magnitudes 
of the field stars, corresponding to $\lambda_{\rm eff} = 7439$~\AA.
 \end{list}
\end{table}

The $u'$-band magnitude of the host was calibrated with observations
of three standard stars in the PG~0231+051 \citet{Landolt92} field
taken the same night with an airmass of 1.0. Assuming an extinction
coefficient in the range 0.36--0.4, the zero point for the $u'$-band
filter was estimated to be $Z = 27.8 \pm 0.3$~mag.
Table~\ref{tab:host_mag} reports the photometry of the host galaxy.

The host centroid, as determined with SExtractor, is $\alpha$(J2000)
$=18^{\rm h} 37^{\rm m} 38\fs03$, $\delta$(J2000) $=+62^\circ
44\arcmin 39\farcs0$, which is $0\farcs4$ away from the afterglow
position. This angular offset corresponds to a projected distance of
3.4~kpc. Taking into account the uncertainty in the afterglow
position, an upper limit of 6~kpc is more conservative, in agreement
with the typical projected offsets of long-duration GRBs
\citep{Bloom02}.

\subsubsection{Spectroscopy}
\label{sec:spectrum}

We initiated observations of the GRB afterglow with the Gemini/GMOS
dual spectrometer starting at time 13:24 for a series of two 1200~s
exposures using the R400 grating, offset by $\sim 5$~\AA\ to fill in
the detector gaps.  The data were reduced with the LowRedux
pipeline\footnote{http://www.ucolick.org/$\sim$xavier/LowRedux/index.html.}
developed by J.X.P. and J. Hennawi.  This custom software bias
subtracts, flatfields, and optimally extracts the spectra.  The two
exposures were coadded, weighting by signal-to-noise ratio (S/N); the
resultant spectrum has a S/N exceeding 80 per pixel over the range
$\lambda \approx 5000$--8000~\AA.

\begin{figure*}
\centering
\includegraphics[width=10cm,angle=90]{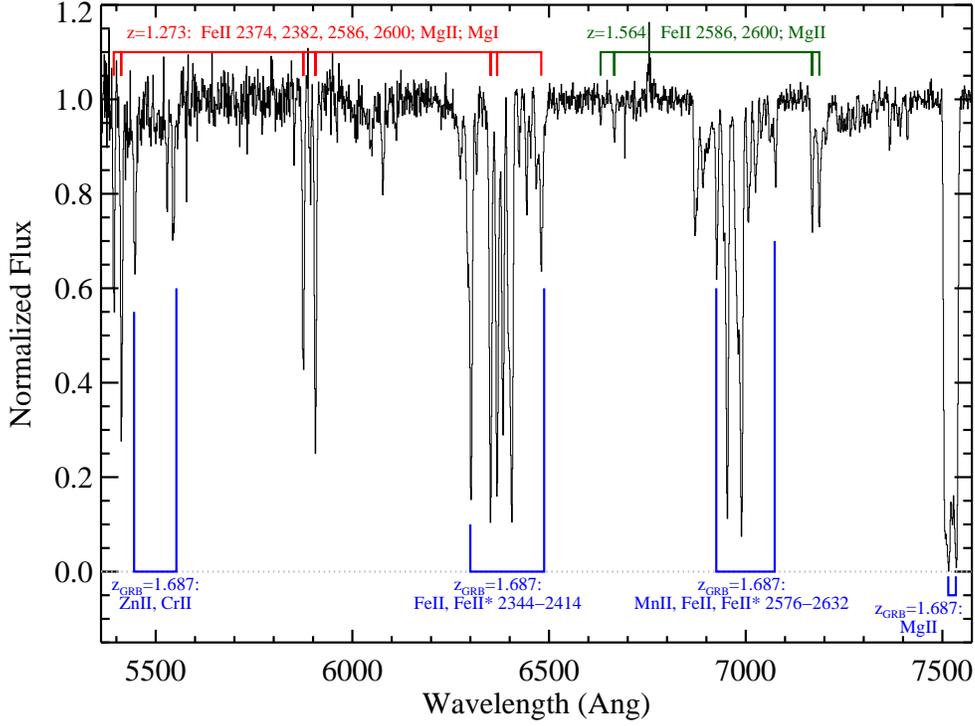}
\caption{Gemini/GMOS spectrum of the afterglow taken at 2.1~hours
  post burst.}
\label{f:gem_spectrum}
\end{figure*}
%

The spectrum reveals a series of very strong rest-frame UV transitions
at $z = 1.68742$ that includes Al~III $\lambda\lambda$1854, 1862,
Zn~II $\lambda$2026, Mg~II, and a plethora of Fe~II resonance and
fine-structure transitions (Fig.~\ref{f:gem_spectrum}).  The detection
of the fine-structure transitions uniquely identifies this gas (and
redshift) with the host galaxy of GRB~080603A \citep{Prochaska06}.  We
have also detected two intervening Mg~II systems along the sightline
at $z = 1.2714$ and $z = 1.5636$. The former marks yet another example
of a strong (equivalent width $W_{2796} > 2$~\AA) Mg~II system along a
GRB sightline \citep{Prochter06}.  Table~\ref{tab:ew} lists the
equivalent-width measurements (from Gaussian fits to the absorption
lines) for all of the features detected in our spectrum.  Note that
the total equivalent width is reported for transitions that are
severely blended.

\subsection{Radio data}
\label{sec:radio}

GRB~080603A was observed with the VLA\footnote{The NRAO is a facility
  of the National Science Foundation (NSF), operated under cooperative
  agreement by Associated Universities, Inc.} at four epochs in the
8.5~GHz band and at 2 epochs in the 4.9~GHz band. Our observations
spanned from 2008 June 5 until June 16. That on June 5 was made in the
VLA C-configuration, whereas the later observations were made in the
DnC-configuration.  We adopted the VLA calibrator J1835+613 for phase
calibration at both frequency bands.

The data were analysed using standard data-reduction routines of the
Astronomical Image Processing System (AIPS). 3C~286 was used for the
flux calibration.  We had a possible detection ($\sim 2.8\sigma$) in
the first observation at 8.5~GHz; it was confirmed in the subsequent
observations.  The results are reported in Table~\ref{tab:radio}.

\begin{table}
\centering
  \caption{Flux densities of the radio afterglow obtained with the VLA.}
  \label{tab:radio}
  \begin{tabular}{lrcr}
\hline
UT Date & $\Delta t$ & $\nu$ & $F_{\nu}$\\
        &  (days)     & (GHz) & ($\mu$Jy)\\
\hline
June 05.39 &  $1.92$ & $8.46$ & $116 \pm 41$\\
June 07.42 &  $3.95$ & $8.46$ & $154 \pm 28$\\
June 08.24 &  $4.77$ & $4.86$ & $112 \pm 45$\\
June 08.26 &  $4.79$ & $8.46$ & $230 \pm 29$\\
June 16.19 & $12.72$ & $4.86$ & $186 \pm 49$\\
June 16.21 & $12.74$ & $8.46$ & $ 70 \pm 42$\\
\hline
\end{tabular}
\end{table}
%

\section{Results}
\label{sec:res}

\subsection{Multi-band light curves}
\label{sec:lc}

\begin{figure*}
\centering
\includegraphics[width=15cm]{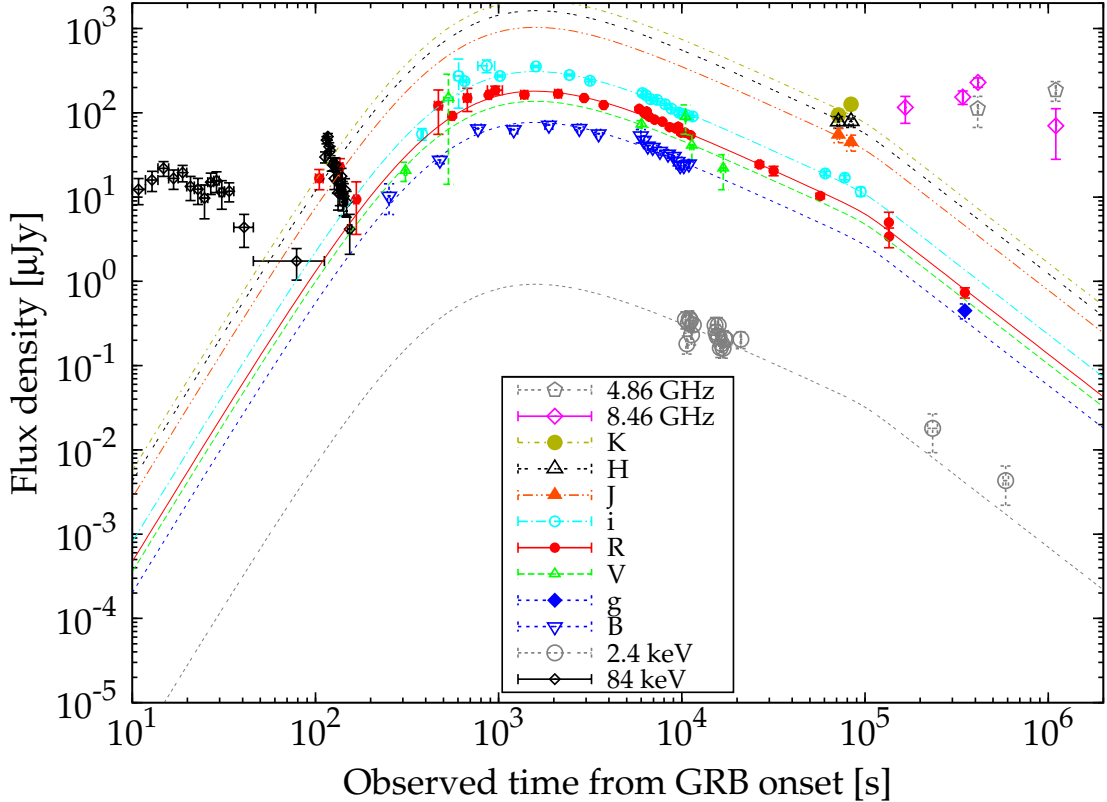}
\caption{Panchromatic light curves of the prompt and afterglow emissions.}
\label{f:lc_panchro}
\end{figure*}

Figure~\ref{f:lc_panchro} shows the $\gamma$-ray prompt and broadband
afterglow light curves. 
In modelling the data we initially allowed colour change between
$5\times10^3$ and $10^5$~s, i.e. the best--sampled interval, by fitting the
data with the sum of a fast-decaying component and a slow-decay component.
The colour change between the two was $0.225\pm0.344$, so less than 1-$\sigma$.
Depending on whether or not we allow colour change, the results do not change
within uncertainties. Given the apparent lack of chromatic changes,
to better constrain the evolution we simultaneously fitted the light curves of the various
bands (except at the radio wavelengths) with the same function, only
allowing different normalisations and no colour change.  We modelled the
different power-law regimes with the smoothly broken power-law model
parametrisation by \citet{Beuermann99}. Furthermore, given the clear
presence of a break in the $R/r'$ and $B/g'$ curves around $10^5$~s,
we allowed a further achromatic break. The final model is described by
Equation~(\ref{eq:multifit}).
%
%
\begin{equation}
\displaystyle F(t)\,= F_0\ \frac{\left[1 + \left(\frac{t}{t_{\rm
        b2}}\right)^{n_2}\right]^{(\alpha_2-\alpha_3)/n_2}}{\left[\left(\frac{t}{t_{\rm
        b1}}\right)^{n\,\alpha_1} + \left(\frac{t}{t_{\rm
        b1}}\right)^{n\,\alpha_2}\right]^{1/n}}
\label{eq:multifit}
\end{equation}
The best-fitting solution was found through minimisation of the
overall $\chi^2$, resulting from the sum of the total $\chi^2$ values
of the individual light curves with respect to each corresponding
model. The free parameters are the following: $\alpha_1$, $\alpha_2$,
and $\alpha_3$ are the power-law indices during the initial rise, the
following decay, and the final ($>10^5$~s) decay, respectively.  The
two break times are $t_{\rm b1}$ and $t_{\rm b2}$, while $n$ and $n_2$
are the smoothness parameters regarding the first and second breaks,
respectively. Only $n_2$ could not be determined from the fit because
of the sparseness of the data around the final break and it was
therefore fixed to 10, so as to give a rather sharp break. The
normalisation term is represented with $F_0$ in
Equation~(\ref{eq:multifit}), although in practice the free parameter
we used for each profile was the flux density calculated at a fixed
reference time (we chose $1.5 \times 10^4$~s, close to the time of
most observations).  The purpose of this choice is to limit the
effects of the strong correlations between some parameters, such as
$\alpha_1$ and $t_{\rm b1}$, on the determination of the confidence
intervals for each parameter.

Concerning the late-time $g'$ point, we converted its flux density
into that corresponding to the close-in-frequency $B$ band by the
factor $f = (\nu_B/\nu_g)^{-\beta_{\rm obs}} \approx 0.9$, where
$\beta_{\rm obs} = 2$ is the observed spectral index of the SED at the
observed visible wavelengths (Section \ref{sec:sed}). This allows us to
better constrain the late-time break, in addition to that offered by
the $R$-band curve.  The same correction was applied to the few $r'$
flux-density points to shift them into the $R$ band.

This correction introduces a negligible systematic uncertainty.
We verified this by alternatively fitting a SED derived by treating all
the filters separately, and the best-fitting parameters and corresponding
uncertainties turned out to fully agree with the values obtained
by merging the filters above mentioned and presented in Section~\ref{sec:sed}
(see also \citealt{Kann11}).

The best-fitting models are shown in Figure~\ref{f:lc_panchro} and the
corresponding best-fitting parameters are reported in
Table~\ref{tab:multifit}, where the normalisation terms are expressed
as flux densities calculated at $1.5 \times 10^4$~s.

\begin{table}
\centering
  \caption{Multi-filter light curve best-fitting parameters.}
  \label{tab:multifit}
  \begin{tabular}{lcc}
\hline
Parameter & Value & Unit\\
\hline
$\alpha_1$           &  $-3.6 \pm 1.7$ & \\
$t_{\rm b1}$         &  $610_{-251}^{+470}$ & s\\
$\alpha_2$           &  $0.99 \pm 0.07$ & \\
$n$                  &  $0.30_{-0.13}^{+0.23}$ & \\
$n_K$          &  $505_{-113}^{+145}$ & $\mu$Jy\\
$n_H$          &  $378_{-77}^{+97}$ & $\mu$Jy\\
$n_J$          &  $240_{-60}^{+81}$ & $\mu$Jy\\
$n_i$          &  $73 \pm 3$ & $\mu$Jy\\
$n_r$          &  $42.0 \pm 1.5$ & $\mu$Jy\\
$n_V$          &  $32.2_{-5.0}^{+6.0}$ & $\mu$Jy\\
$n_B$          &  $18.0 \pm 0.9$ & $\mu$Jy\\
$n_X$          &  $0.22 \pm 0.03$ & $\mu$Jy\\
$t_{\rm b2}$         &  $1.0_{-0.4}^{+0.6}$ & $10^5$~s\\
$\alpha_3$           &  $1.7_{-0.3}^{+0.4}$ & \\
$\chi^2/{\rm dof}$   &  $102.4/86$\\
\hline
\end{tabular}
\end{table}

The peak time, $t_{\rm p}$, is a function of the free parameters as
expressed by Equation~(\ref{eq:tp}):
\begin{equation}
t_{\rm p}\ =\ t_{\rm b1}\ \Big(\frac{-\alpha_1}{\alpha_2}\Big)^{1/n
  (\alpha_2-\alpha_1)}\ =\ 1575_{-250}^{+430}\ \textrm{s}
\label{eq:tp}
\end{equation}
The uncertainty in $t_{\rm p}$ was calculated through error
propagation, taking into account the covariance of parameters.
Equation~(\ref{eq:tp}) is exact only when no further breaks are
present --- that is, when $t_{\rm b2} = \infty$ (or, equivalently,
$\alpha_3 = \alpha_2$).  In practice, it still holds provided that
$t_{\rm b2} \gg t_{\rm p}$, as in this case.

From the fit we excluded the earliest points at $t < 155$~s, connected
with the optical flash observed contemporaneously with the last
$\gamma$-ray pulse (see Fig.~\ref{f:lc_panchro}). We accounted for the
presence of some degree of variability around the best-sampled curves,
particularly around the broad peak at $t_{\rm p}$, by adding in
quadrature a systematic error to the statistical ones (7\%, 4\%, and
8\% for the $B$, $V$, and $i'$ curves, respectively), so as to yield a
satisfactory goodness of the fit: $\chi^2/{\rm dof}$ = 102.4/86
($p$-value of 11\%). The reason for the additional errors was to avoid
the risk of underestimating the fit parameters' uncertainties. We also
note that the nature of this additional systematic scatter cannot be
entirely ascribed to unaccounted variability of the zero points; at
least a few percent must genuinely characterise the afterglow light
curve.  Indeed, the largest deviations from the models are seen to
occur {\em simultaneously} and to {\em correlate} in all of the more
densely sampled curves: $B$, $R$, and $i'$.  Such deviations are
mostly observed within the interval 500--1000~s (final part of the
rise) and around $10^4$~s, at the beginning of the decay.

We alternatively adopted a different log-likelihood function, one
more general than that connected with the $\chi^2$ of equation~(\ref{eq:multifit})
in that it treats the additional systematic errors as free parameters.
In practice, this approach does not provide any noticeable difference
in the best-fitting parameters and uncertainties, and the physical
implications discussed in Section~\ref{sec:disc} are completely unaffected. 
The only difference concerns the systematic error affecting the B--filter
values, for which the alternative log-likelihood provides a systematic error
compatible with zero within uncertainties. In any case, this does not
affect the best-fitting model to any noticeable degree.

Admittedly, because of the paucity of NIR points (only two, very close
in time in each filter), the fitting models in
Figure~\ref{f:lc_panchro} at the corresponding wavelengths assume an
achromatic evolution extended to the NIR bands.  At first glance this
may seem too arbitrary, as there have been GRBs, particularly those
with SEDS that are temporally well resolved, for which some chromatic
evolution was observed, such as GRB~061126 \citep{Perley08c},
GRB~071025 \citep{Perley10}, and GRB~080319B
\citep{Bloom09,Racusin08}. Nonetheless, we note that no chromatic
evolution was required by comparably early-time observations of
several other GRBs (e.g.,
\citealt{Kruhler09,Nysewander09a,Yuan10,Covino10,Perley11}).

Such an assumption has also been made implicitly for the X-ray data;
these are too sparse to be modelled independently. Our results show
that the X-ray light curve can be described with the same rigid model
as the NIR/optical, but of course other possibilities cannot be ruled
out.  Furthermore, the two latest X-ray points in
Figure~\ref{f:lc_panchro} assume the same spectral shape as that
observed around $1.5 \times 10^4$ due to paucity of X-ray photons
observed after $10^5$~s (Section \ref{sec:xray}). Although the X-ray photon
index of 2.3 observed at $10^4$~s is not expected to significantly
evolve, the assumption of no X-ray spectral evolution from $10^4$~s to
$10^5$~s must be pointed out.

The radio data clearly show a different behaviour: the peak of the
afterglow spectrum crosses the radio bands a few days later. Because
of this, they were not considered in the achromatic modelling, but
deserve a dedicated analysis in the framework of the standard
afterglow model discussed in Section \ref{sec:disc:model}.

\subsection{Spectral energy distribution}
\label{sec:sed}

\begin{figure}
\centering
\includegraphics[width=8.5cm]{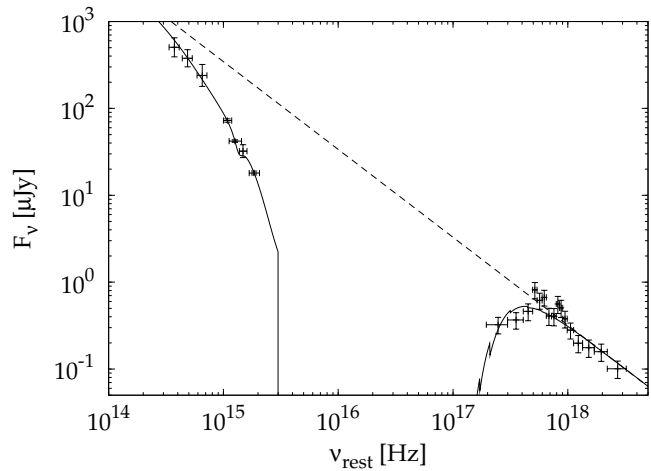}

\caption{Rest-frame SED at $1.51 \times 10^4$~s.  The solid line shows
  the best-fitting power law with a LMC2 dust-extinction
  profile. The dashed line shows the same unextinguished power law.}

\label{f:sed}
\end{figure}

We derived an SED from the multi-filter light-curve fitting. Although
this is based on the achromatic evolution and, as such, does not refer
to a particular epoch, we considered the reference time $t_{\rm ref} =
1.5 \times 10^4$~s as the most representative of it: this is the
midpoint time of the X-ray data, when the high-energy end of the SED
does not rely on any assumption. Figure~\ref{f:sed} displays the
resulting GRB rest-frame SED.
%
%

We considered two different models, either a simple power-law and a
broken power law with a cooling break, $\Delta$$\beta=0.5$, combined
with three different extinction profiles according to the
parametrisation of \citet{Pei92}: SMC, LMC2 (for the Large Magellanic
Cloud), and Milky Way (MW).  Note that while Pei's measurements of the
average MW and SMC extinction curves are generally consistent with
recent estimates (e.g., \citealt{Gordon03}), the LMC implementation is
representative only of the area around 30~Doradus (the LMC2
supershell); hence, we denote this curve as LMC2 in this work. The
average LMC curve is much more similar to that of the Milky Way.

Table~\ref{tab:sed} reports all of the results of fitting the SED with
different extinction profiles and/or different models at different
energy ranges. The uncertainties on the best-fitting parameters
include the dependence of the effective frequency of each filter on the folded model.
Interestingly, only the LMC2 profile can satisfactorily
account for the NIR/optical SED, possibly including the presence of the
2175\,\AA\ bump.  The dust content is remarkable: the rest--frame
extinction is $A_{V,z} = 0.80 \pm 0.13$ mag, one of the highest among
GRBs having observed optical afterglows \citep{Kann10}.
The other two models, MW and SMC, are ruled out.
Concerning the models, a simple power law from NIR to
X-rays provides a very good result, with $\beta_{\rm ox} = 1.01 \pm
0.05$.

The quality of the photometric set derived for GRB~080603A allows us
to attempt a full parametric dust characterisation using the general
parametrisation of the Local Group extinction laws by
\citet{Fitzpatrick90} (hereafter FM), like with GRB~080607
\citep{Perley11}.  In our analysis the parameters $c_1$, $c_2$, and $R_V$
were all tied to each other, as for GRB~080607.
The other parameters that were fixed are $\gamma$ to 1 and $c_4$ to 0.6,
respectively accounting for the 2175\,\AA\ bump
width and for the strength of the far-UV rise.  The free parameters
were $\beta = 0.98 \pm 0.04$, $A_{V,z} = 0.57 \pm 0.19$, $R_V = 2.14
\pm 0.33$, and $c_3 = 1.76 \pm 0.66$ (strength of the 2175\,\AA\ bump)
with $\chi^2/{\rm dof} = 19.8/18$.  Indeed, comparing with the
corresponding values for the LMC2 supershell, $R_V = 2.76 \pm 0.09$
and $c_3 = 1.46 \pm 0.12$ \citep{Gordon03}, confirms that the FM model
converges to values generally consistent with those of the LMC2
profile. In particular, the 3$\sigma$ nonzero value of $c_3$ shows
that the 2175\,\AA\ bump is likely to be present.

As a further check, we folded the
spectral models with the filters' transmission curves and iteratively
found the new effective frequencies at which the model flux densities
were the same as the folded ones. We recalculated the best--fit model
based on the new effective frequencies. This sequence was repeated
until the best-fitting parameters converged to a stable solution.
The final effective frequencies changed by 5\% at most from the
corresponding nominal values which had not been folded with the
transmission curves.
The best-fitting parameters and the possible evidence for the
2175\,\AA\ bump were confirmed with the same confidence.

Although a broken power-law model cannot be ruled out, in this case
the break frequency must lie within the soft end of the X-ray band. We
therefore conclude that the break frequency (if any) must lie either
outside the optical to X-ray range or within the X-ray band itself.
\begin{table*}
\centering
  \caption{Spectral energy distribution best-fitting parameters.}
  \label{tab:sed}
  \begin{tabular}{clccccccc}
\hline
Frequency Range & Model & Ext. profile & $\beta$           & $A_{V,z}$ & $N_{{\rm H~I},z}$ & $\nu_{\rm b}$ & $\chi^2/{\rm dof}$ & Prob\\
($10^{15}$~Hz)   &        &       &                   &               & ($10^{21}$~cm$^{-2}$) & ($10^{17}$~Hz) &  & (\%) \\
\hline
0.3--$3 \times 10^3$ & {\sc pow} & LMC2    & $1.01\pm0.05$    &  $0.80\pm0.13$ & $6.7_{-1.8}^{+2.0}$ & --       & $19.0/19$ & $46$\\
0.3--$3 \times 10^3$ & {\sc pow} & SMC     & $0.92\pm0.04$    &  $0.48\pm0.07$ & $5.8_{-1.6}^{+1.8}$ & --       & $39.9/19$ & $0.34$\\
0.3--$3 \times 10^3$ & {\sc pow} & MW      & $1.04\pm0.04$    &  $0.91\pm0.12$ & $6.5_{-1.6}^{+1.9}$ & --       & $136/19$  & $<10^{-15}$\\
0.3--3           & {\sc pow} & LMC2    & $0.8\pm0.7$      &  $0.9_{-0.3}^{+0.4}$ & --            & --       & $0.4/4$   & $98$\\
0.3--3           & {\sc pow} & SMC     & $2.0_{-0.6}^{+0.5}$ &  $<0.33$       & --               & --       & $12.1/4$  & $1.7$\\
0.3--3           & {\sc pow} & MW      & $2.2_{-0.4}^{+0.2}$ &  $<0.28$       & --               & --       & $10.5/4$  & $3.3$\\
200--$3 \times 10^3$ & {\sc pow} & --      & $1.3\pm0.3$ &  -- & $6.6_{-4.6}^{+6.2}$                 & --       & $16.4/13$ & $23$\\
\\
0.3--$3 \times 10^3$ & {\sc bkn} & LMC2    & $0.99\pm0.07$    &  $0.80\pm0.13$ & $8.0_{-3.1}^{+3.2}$ & $8.7_{-2.7}^{+4.2}$ & $16.6/18$ & $55$\\
0.3--$3 \times 10^3$ & {\sc bkn} & SMC     & $0.89\pm0.05$    &  $0.48\pm0.07$ & $7.1_{-1.8}^{+3.1}$ & $8.2_{-4.0}^{+3.8}$ & $34.9/18$ & $1.0$\\
\hline
\end{tabular}
\end{table*}

Figure~\ref{f:sedradio} shows a SED at $4.1 \times 10^5$~s including
two radio-flux measurements.  Under the assumption that there is no
break frequency between radio and visible filters apart from the peak
synchrotron frequency $\nu_{\rm m}$, we tried to model the radio
points as lying in the $F_\nu \approx \nu^{1/3}$ power-law segment for
$\nu < \nu_{\rm m}$ in the slow-cooling regime \citep{Sari98} with the
cooling frequency lying above the X-rays.  We extrapolated the optical
flux densities of $R$ and $g'$, the only filters with observations
taken at a comparable epoch, assuming the temporal decay of Section
\ref{sec:lc}. The X-ray spectrum was also rescaled accordingly.  We
assumed a negligible contribution from the reverse shock.  This would
allow us to constrain $\nu_{\rm m}$ at the same time: it turns out to
be $\nu_{\rm m}\,(1+z) = (4.4 \pm 0.8) \times 10^{12}$~Hz in the GRB
rest frame (dashed-dotted line in Fig.~\ref{f:sedradio}).  However, as
will be shown in Section \ref{sec:disc:radio}, this clashes with the
observed temporal evolution of the afterglow, particularly in the
radio band: from the light curve at 8.46~GHz $\nu_{\rm m}$ must have
crossed the radio band immediately thereafter, around $4.4 \times
10^5$~s. This would imply an unreasonably fast decay of $\nu_{\rm m}$.
In addition, the resulting peak flux of 1.2~mJy is much higher than
that observed in the radio when $\nu_{\rm m}$ crossed it.

\subsection{Optical flash}
\label{sec:flash}

Figure~\ref{f:lc_panchro} clearly shows that the flux densities at
optical and $\gamma$-ray wavelengths during the optical flash that
occurred simultaneously with the last $\gamma$-ray pulse are nearly
equal.  The corresponding average spectral index is therefore
$\beta_{{\rm opt} - \gamma} \approx 0$.  Given the considerable amount
of dust, the dust-corrected optical flux increases by a factor of
$\sim 7$, obtained by the LMC2 extinction profile that best fits the
SED at the GRB rest-frame frequency of $1.3 \times 10^{15}$~Hz,
corresponding to the observed $R$ filter. We point out that the dust
considered here is that within the host galaxy, since the Galactic
term had already been removed.  Replacing the observed optical flux
density with the dust-corrected value, the average spectral index
becomes $\beta_{{\rm opt}-\gamma} = 0.13$. The $\gamma$-ray spectral
index during the last pulse is $\beta_\gamma = 0.65 \pm 0.2$
(Table~\ref{tab:gamma}).  

On the one hand, a simple extrapolation of the $\gamma$-ray spectrum
to optical wavelengths overpredicts the dust-corrected optical flux by
$2 \pm 1$ orders of magnitude; this has already been observed for
other bursts with optical detections during the prompt emission
(\citealt{Yost07}; note that these authors adopted a different
convention for the sign of $\beta$).  On the other hand, given the
intermediate value of $\beta_\gamma$ lying between the most common
values of 0 and 1.3 expected for the low-energy and high-energy
indices (respectively) of a typical Band function, $E_{\rm p}$ could
lie either within or close to the 20--200~keV band with a low-energy
index close to 0, as we argued in Section \ref{sec:gamma}.  In this
context, the dust-corrected $\beta_{{\rm opt}-\gamma}$ value is fully
consistent with the low-energy index distribution observed in the
prompt spectra of most GRBs \citep{Kaneko06} and the optical flux
density would match the extrapolation of the prompt $\gamma$-ray
spectrum. However, a broadband flat spectrum and a correspondingly
flat electron energy distribution is somewhat nonstandard in the
synchrotron shock model.

An interesting possibility proposed in the literature interprets the
optical flash as the result of internal shocks with lower velocity
irregularities at larger radii, as suggested for the flash of
GRB~990123 \citep{Meszaros99}.

Temporal analysis of both profiles adds little information: because of
the relatively coarse optical sampling of FTN, a correlation between
the optical and $\gamma$-ray fluxes is neither confirmed nor ruled
out. This was tested by integrating the $\gamma$-ray counts in the
time windows of the optical frames and by comparing the relative
variations between the two bands.  Although the three optical points
exhibit the same behaviour as the $\gamma$-rays, both cases
(correlation or lack thereof) are compatible with the data within
uncertainties.

Although the measured optical flux can be the low-energy extrapolation
of the prompt $\gamma$-ray emission, in Section \ref{sec:disc:flash} we
test the possibility that the $\gamma$-rays are upscattered photons of
the optical flash, as suggested for other GRBs to overcome the
problems of the synchrotron model (e.g., \citealt{Kumar08},
\citealt{Racusin08}).

\subsection{Host galaxy}
\label{sec:sed_host}

\begin{figure}
\centering
\includegraphics[width=8.5cm]{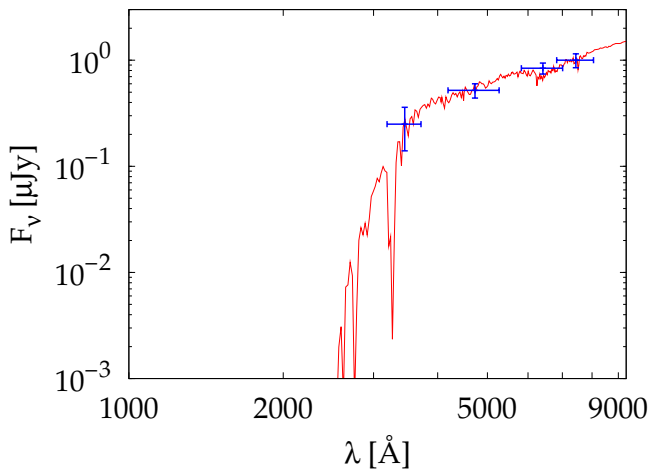}
\caption{Observed SED of the host galaxy and the best-fitting
  synthetic spectrum using {\tt hyperZ}.}
\label{f:sed_host}
\end{figure}

We compared the observed host-galaxy SED (Table~\ref{tab:host_mag})
with a set of spectral synthesis models by using {\tt hyperZ} (see
\citealt{Bolzonella00} for details).  The lack of measurements redward
of the Balmer break at 4000~\AA (rest frame) limits our possibility of
constraining one of the key parameters, the total stellar mass.  We
assumed different extinction profiles for modelling the dust
extinction produced within the galaxy itself; although the best-fit
parameters do not change significantly, we adopted the Calzetti
law \citep{Calzetti00}. The result is shown in Figure~\ref{f:sed_host}.

%
%

The best-fitting synthetic spectrum belongs to a starburst galaxy with
an age of 130~Myr, dust extinction $A_V = 0.87$~mag, absolute
magnitude $M_B = -20.7$. By adopting an LMC extinction profile,
similar parameters are obtained: 90~Myr age, $A_V = 0.77$~mag, $M_B =
-20.7$.  These values are typical of other host galaxies of
long-duration GRBs \citep{Savaglio09,Chen09}.

\section{Discussion}
\label{sec:disc}

GRB~080603A exhibits several interesting properties: (i) the end of the
$\gamma$-ray prompt emission marked by the simultaneous detection of
an optical flash, which appears to be a distinct component from the
emerging afterglow following the end of the $\gamma$-rays; (ii) an
overall achromatic afterglow rise, peak, and decay, followed by a break
around $10^5$~s; (iii) an accurate SED allowing us to precisely measure
the remarkable dust content along the sightline to the GRB, clearly
favouring an LMC2 profile at variance with that found for most GRBs.
In the following we examine these aspects in the context of the
standard afterglow model.

\subsection{Broadband afterglow modelling}
\label{sec:disc:model}

In the context of the standard afterglow model
(\citealt{Meszaros97,Sari98}; for a review, see, e.g.,
\citealt{Meszaros06}), the power-law piecewise spectra and light
curves are interpreted as the result of synchrotron emission of a
population of shock-accelerated electrons of a forward shock ploughing
into the surrounding medium. The electron energy distribution is
assumed to be $dN/d\gamma \propto \gamma^{-p}$ ($\gamma > \gamma_m$);
the values of $p$ typically derived from GRB afterglow modelling
cluster around 2.0--2.4 with a scatter of 0.3--0.5
\citep{Starling08,Curran10}, in general agreement with theoretical
expectations (e.g., \citealt{Achterberg01,Spitkovsky08}).

Following \citet{Sari98}, let $\nu_{\rm m}$ and $\nu_{\rm c}$ be the
synchrotron injection and cooling frequencies, respectively. The
observed spectral index $\beta_{\rm ox} = 1.0$ (Section \ref{sec:sed}) with
no breaks between optical ($\nu_{\rm o}$) and X-ray ($\nu_{\rm x}$)
frequencies can be explained in two alternative cases: $\nu_{\rm m} <
\nu_{\rm o,x} < \nu_{\rm c}$ or ${\rm max}(\nu_{\rm m},\nu_{\rm c}) <
\nu_{\rm o,x}$. In the latter case, the electron index is $p =
2\,\beta_{\rm ox} = 2.0$ (although formally consistent with fast
cooling, at late times we can safely assume slow cooling), while in
the former case $p = 2\,\beta_{\rm ox} + 1 = 3.0$ (slow cooling).
%
\begin{itemize}

\item $\nu_{\rm m} < \nu_{\rm o,x} < \nu_{\rm c}$ ($p=3$).

The predicted temporal index $\alpha$ depends on the density profile:
either $\alpha_{\rm ISM} = 3(p-1)/4 = 1.5$ (homogeneous or ISM) or
$\alpha_{\rm w} = (3p-1)/4 = 2.0$ (wind). Both decay values are
significantly steeper than the observed $\alpha_2 = 1.0$,
respectively, by $\Delta\alpha_{\rm ISM} = 0.5$ and $\Delta\alpha_{w}
= 1.0$.  Even assuming a more general density profile, $n(r) \propto
r^{-s}$, the expected decay is steeper than ISM for every value of
$s$.  This was also the case for a number of other bursts whose
spectral and temporal indices could not fulfil any closure relation
(e.g., \citealt{Melandri08}).  A possible way to explain a shallower
decay is energy injection refreshing the blast wave
\citep{Rees98,Sari00,Granot03,Melandri09}, as was also proposed to
explain the shallow-decay phase in early X-ray afterglows
\citep{Zhang06}.

Let $E(t)$ be the fireball energy as a function of the observed time
$t$, so that $E(t) \propto t^e$. Assuming negligible radiative losses,
the expected decay index change with respect to no injection,
$\Delta\alpha_{ei}$, is $e(p+3)/4$ (ISM) and $e(p+1)/4$ (wind) for
$\nu < \nu_{\rm c}$ \citep{Panaitescu05}. The observed values imply $e
= 1/3$ (ISM) and $e = 1$ (wind).  The energy budget may be
problematic, as in the afterglow phase from $10^3$ to $10^5$~s the
required injected energy would be a factor of $100^{1/3} \approx 5$
(100) larger for ISM (wind).  Using the notation of \citet{Zhang06},
these are equivalent to $q = 2/3$ and $q = 0$, respectively, for an
injection luminosity $L(t) \propto t^{-q}$.

At the end of the injection, the power-law decay is expected to
steepen by $\Delta\alpha_{ei}$ so that the decay should resume the
no-injection values, $\alpha_{\rm ISM}$ or $\alpha_{\rm w}$. Indeed,
the final decay $\alpha_3 = 1.7_{-0.3}^{+0.4}$ is compatible with both
values. In this case, the final break around $10^5$~s would mark the
end of the energy-injection process.

\item ${\rm max}(\nu_{\rm m},\nu_{\rm c}) < \nu_{\rm o,x}$ ($p=2$).

Above the cooling frequency the emission does not depend on the
ambient density, so for any $s$ ($0 < s < 3$), $\alpha = (3p-2)/4 =
1.0$. This fully agrees with the observed value of $\alpha_2$ and
there is no need to invoke any additional processes such as energy
injection. Another asset of this possibility is that the final
power-law decay index, $\alpha_3 \approx 2$, nicely supports the
jet-break interpretation, being $\alpha = p$.
\end{itemize}

\subsubsection{Radio afterglow}
\label{sec:disc:radio}

The peak frequency $\nu_{\rm m}$ of the synchrotron spectrum crossed
the radio bands between 5 and 12~days: this is suggested by the peak
in the light curve at $\nu_{\rm radio,2} = 8.46$~GHz and by the change
in the spectral slope between $\nu_{\rm radio,1}=4.86$~GHz and
$\nu_{\rm radio,2}$.  In particular, the slope of the radio spectrum
changes from positive to negative: the first 2-channel radio SED at
5~days is fit with $\beta_{\rm radio} = -1.3_{-2.0}^{+1.1}$, while the
second SED around 12~days gives $\beta_{\rm radio} =
1.8_{-1.6}^{+7.2}$.  At low frequencies the flux is expected to rise
as $t^{1/2}$ and to decay as $t^{-3(p-1)/4}$ for an ISM
\citep{Sari98}. We fitted the observed 8.46~GHz radio curve under
these assumptions in either case considered above (i.e., either $p=3$
or $p=2$) and found the radio peak time $t_{\rm radio,p} =
4.4_{-0.7}^{+3.7} \times 10^5$~s and the peak normalisation $F_{\rm
  radio,p} = 200_{-40}^{+120}$~$\mu$Jy from the fit. This could be
considerably different, when other temporal behaviours are considered;
in particular, a steeper rise followed by a steeper decay could be
compatible with $F_{\rm radio,p} > 200$~$\mu$Jy. This possibility is
not naturally explained within the standard afterglow model, unless
the $\nu_{\rm m}$ passage and the jet break happened almost at the
same time by chance, implying a $t^{-p}$ decay.

\subsubsection{Afterglow onset}
\label{sec:disc:ag_onset}

The rise experienced by the afterglow in the $BVRi$ filters rules out
the passage of a typical synchrotron frequency through the observed
wavelengths as the possible cause because of both the steepness and
the lack of chromatic evolution of the rise itself.  The possibility
of an afterglow emerging from a wind surrounding a massive progenitor,
with the optical rise being due to the progressively decreasing dust
extinction \citep{Rykoff04}, is excluded by the lack of chromatic
evolution, as observed in many other cases (e.g.,
\citealt{Guidorzi09,Kruhler09}).

Here we consider the possibility, discussed in several analogous cases
(e.g., \citealt{Molinari07}), that the broad optical peak $t_{\rm p} =
1575_{-250}^{+430}$~s marks the deceleration of the fireball and the
afterglow onset.  The duration of the $\gamma$-ray burst itself ($\sim
200$~s) is much shorter than the peak time, as expected in the
thin-shell case \citep{Sari97}.  The observed steep rise, $\alpha_1 =
-3.6 \pm 1.7$, rules out the wind environment for which a shallower
rise ($\sim -0.5$) is required. Depending on whether it is $\nu <
\nu_{\rm c}$ or $\nu > \nu_{\rm c}$, a rise index of $-3$ or $-2$ is
expected for an ISM \citep{Jin07,Panaitescu08}, both compatible with
observations.

In this context, from the afterglow peak time we can estimate the
initial bulk Lorentz factor $\Gamma_0$ as being approximately twice as
large as its value at the peak time \citep{Sari99,Molinari07}:
\begin{equation}
\Gamma_0\ \approx\ 2\ \Gamma(t_{\rm p})\ =\ 2\ \Big[\frac{3 E_{\rm
      iso} (1+z)^3}{32 \pi n m_p c^5 \eta t_{\rm
      p}^3}\Big]^{1/8}\ \approx\ (130 \pm 20) n_0^{-1/8}
\label{eq:gamma_ism}
\end{equation}
We assumed standard values for the energy-conversion efficiency,
$\eta_\gamma = 0.2$ and for the particle density of the circumburst
environment $n = n_0$~cm$^{-3}$.  Equation~(\ref{eq:gamma_ism}) holds
for an ISM environment; we do not consider the wind case, because of
the incompatible steepness of the rise.  Such a bulk Lorentz factor
lies within the distribution found for other GRBs (e.g.,
\citealt{Liang10,Melandri10}).

Following \citet{Zou10}, we set an upper limit to $\Gamma_0$ from the
prompt $\gamma$-ray light curve, thanks to the presence of a quiescent
time between the two pulses.  The idea behind this is that while the
prompt $\gamma$-rays are being produced through internal shocks, the
outermost shell begins sweeping up the surrounding medium; as a
consequence, a forward shock should appear in soft $\gamma$-rays as a
smooth and continuously increasing emission. This constraint is only
suitable for bursts with a relatively short pulse followed by either a
quiescent time or a deep trough, because otherwise the external shock
could have significantly decelerated during the first pulse. From
Figure~\ref{f:lc_panchro}, at 90~s we estimate a 3$\sigma$ upper limit
of 3~$\mu$Jy at the mean energy of 84~keV, which corresponds to
$0.3 \times 10^{-28}$~erg~cm$^{-2}$~s$^{-1}$~Hz$^{-1}$. From their
Equation~(5) we derive the following upper limits to $\Gamma_0$:
\begin{eqnarray}
\Gamma_0 & < & 150\, n_0^{-1/8}\, \epsilon_{e,-1/2}^{-1/5}\,
\epsilon_{B,-1}^{-1/20}\, (1+Y)^{1/10}\quad (p=3)\\ 
\Gamma_0 & < & 220\, n_0^{-1/8}\, \epsilon_{e,-1/2}^{-1/8}\,
(1+Y)^{1/8}\quad\quad\quad (p=2.01),
\end{eqnarray}
where we used $p=2.01$ instead of $p=2$ as it is applicable only for
$p>2$.  A value $p<2$ would imply electron energy divergence, so a
proper formulation is required; this is beyond the scope of the paper.
$\epsilon_{B} = \epsilon_{B,-1} \times 10^{-1}$ and $\epsilon_{e} =
\epsilon_{e,-1/2} \times 10^{-1/2}$ are the equipartition factors for
the magnetic and the electron energy densities, respectively, and $Y$
is the Compton parameter for synchrotron self-Compton scattering.
These upper limits are remarkably consistent with the estimate derived
in Equation~(\ref{eq:gamma_ism}), especially because a number of GRBs
were found to exceed these values.

In the context of peaks interpreted in terms of outflow deceleration
and afterglow onset, \citet{Panaitescu11} proposed that fast-rising
optical afterglows are likely caused by an impulsive ejecta release
with a narrow distribution of Lorentz factors after the GRB itself, in
contrast to an extended release or a broad range of Lorentz factors
more suitable to explain the slow-rising/plateau afterglows.  The
motivation of this interpretation resides in the different
correlations between peak time and peak flux found for each class:
$F_{\rm p} \propto t_{\rm p}^{-3}$ and $F_{\rm p} \propto t_{\rm
  p}^{-1/2}$ for fast-rising and plateau afterglows, respectively.
Following their guideline, when we move GRB~080603A to a common
redshift of $z=2$, its dust-corrected $R$-band peak flux is 0.8~mJy at
a peak time of 1750~s (at $z=2$) --- that is, it lies in the region of
the $F_{\rm p}$--$t_{\rm p}$ plane where the two correlations cross
each other (see Fig.~1 of \citealt{Panaitescu11}). Because of the
$\sim t^3$ rise, GRB~080603A belongs to the peaky afterglow
class. Interestingly, $E_{\rm iso}$ and $F_{\rm p}$ of peaky GRBs are
shown to correlate more tightly than those of plateau GRBs; indeed
GRB~080603A lies very close to the best-fitting power-law relation
shown in Fig.~2 of \citet{Panaitescu11}, in agreement with its being a
member of the peaky class.  The tighter connection between the
$\gamma$-ray released energy and the afterglow peak flux for the
fast-rising GRBs may support the impulsive ejecta release
interpretation.

In the afterglow the presence of a single peak followed by a $\sim
t^{-1}$ decay qualifies GRB~080603A as a Type III member according to
the classification by \citet{Zhang03} and \citet{Jin07}.  At variance
with pre-{\it Swift} expectations mainly based on the case of
GRB~990123 \citep{Akerlof99}, most GRBs show no evidence for a
short-lived reverse-shock peak at early times
\citep{Mundell07,Oates09,Rykoff09,Kann10,Melandri10} and GRB~080603A
is no exception in this respect: the prompt optical flash cannot be
reverse-shock emission because of the gap between the flash and the
afterglow onset.  A way to circumvent this problem is that the outer
layer of the outflow has much higher Lorentz factor compared to the
bulk part of the flow; the outer layer might produce an optical
reverse-shock emission well before the onset of the afterglow, which
is determined by the main part of the flow. However, the rather narrow
spike seems to disfavour this model.  Among the possible explanations,
such as either a high or a low magnetic energy density in the ejecta
(e.g., \citealt{Gomboc08}), here we consider the low-frequency model
\citep{Melandri10}: at the shock-crossing time, marked by the peak,
both injection frequencies of forward and reverse shocks, $\nu_{\rm
  m,f}$ and $\nu_{\rm m,r}$ (respectively), lie below the optical band
\citep{Mundell07}.

Assuming the same microphysical parameters in both shocks, the
relation between the spectral characteristics of the shocks at
$t = t_{\rm p}$ are
\begin{equation}
\frac{\nu_{\rm m,r}}{\nu_{\rm m,f}} \approx \Gamma^{-2},\qquad \nu_{\rm
  c,r} \approx \nu_{\rm c,f},\qquad \frac{F_{\rm max,r}}{F_{\rm
    max,f}} \approx \Gamma,
\label{eq:freq}
\end{equation}
where $F_{\rm max}$ is the peak flux in the frequency domain at a
given time, in this case at $t_{\rm p}$, different from the peak flux
in the time domain at a given frequency, denoted with $F_{\rm p}$.  We
discuss the implications in the two cases considered above.

\begin{table*}
\begin{center}
\begin{tabular}{c||ccc||ccc}
\hline
Shock & \multicolumn{3}{c}{$L \propto t^{-q}$}   &  \multicolumn{3}{c}{$E(>\gamma) \propto \gamma^{-s+1}$}\\
\hline
      & $\nu_{\rm m}$   & $\nu_{\rm c}$  & $F_{\rm max}$  & $\nu_{\rm m}$    & $\nu_{\rm c}$       & $F_{\rm max}$\\ 
\hline
 FS   & $-(q+2)/2$ & $(q-2)/2$  & $(1-q)$      & $-12/(7+s)$ & $-2(s+1)/(7+s)$ & $3(s-1)/2(7+s)$\\ 
 RS   & $-(q+2)/4$ & $(q-2)/2$  & $3(2-3q)/8$  & $-6/(7+s)$  & $-2(s+1)/(7+s)$ & $3(s-2)/2(7+s)$\\ 
\hline
\end{tabular}
\end{center}
\caption{Temporal exponent of injection and cooling frequencies, as
  well as of the maximum flux density for the forward shock (FS) and
  the reverse shock (RS) in the case of continuous energy injection
  through an ISM.  The two formalisms, the luminosity as a function of
  time on the left-hand side, and the energy distribution of the
  shells as a function of the bulk Lorentz factor on the right-hand
  side, are equivalent \citep{Sari00,Zhang06}. The two parameters $q$
  and $s$ are related by $s=(10-7q)/(q+2)$ and $q=(10-2s)/(7+s)$. The
  impulsive case (i.e., no continuous injection) corresponds to $s = q
  = 1$.}
\label{tab:closure_rel} 
\end{table*}

\begin{itemize}
\item $\nu_{\rm m} < \nu_{\rm o,x} < \nu_{\rm c}$ ($p=3$).

This requires $\nu_{\rm m,f} \la \nu_{\rm o}$ and $\nu_{\rm c} >
\nu_{\rm x}$ at $t=t_{\rm p}$, where $\nu_{\rm o} = 5 \times
10^{14}$~Hz and $\nu_{\rm x} = 10^{18}$~Hz.  As discussed above, the
observed temporal decay requires energy injection with $q=2/3$ up to
the late-time break. Temporal evolution of the characteristic
frequencies and peak of the forward shock is $\nu_{\rm m,f} \propto
t^{-(q+2)/2} = t^{-4/3}$, $\nu_{\rm c} \propto t^{(q-2)/2} =
t^{-2/3}$, and $F_{\rm max,f} \propto t^{1-q} = t^{1/3}$
\citep{Zhang06}.  Interpreting the radio flux at $t_{\rm radio} = 4.1
\times 10^5$~s as mostly due to the forward shock, using the result of
the corresponding SED fitting discussed in Section \ref{sec:sed} and shown
in Figure~\ref{f:sedradio} (dashed-dotted line), we can estimate the
time at which $\nu_{\rm m,f}$ crossed the optical bands: $t_{\rm
  radio}\,(\nu_{\rm o}/\nu_{\rm m,f}(t_{\rm radio}))^{-3/4}\,(t_{\rm
  radio}/t_{\rm b2})^{1/8} = 7 \times 10^3$~s.  This is derived from
the temporal dependence of $\nu_{\rm m,f} \propto t^{-4/3}$ as long as
energy injection goes on (so for $t < t_{\rm b2}$), and $\nu_{\rm m,f}
\propto t^{-3/2}$ (for $t > t_{\rm b2}$).  The absence of any break or
spectral evolution in the multi-filter light curve of
Figure~\ref{f:lc_panchro} rules out any such passage at this time.
Even worse, as discussed in Section \ref{sec:sed}, the derived value of
$\nu_{\rm m,f}$ from Figure~\ref{f:sedradio} is incompatible with its
passage through the radio bands at $t_{\rm radio,p}$ as observed from
the radio curve (Section \ref{sec:disc:radio}), because it implies a too
rapid decay of $\nu_{\rm m,f}$.
%
\begin{figure}
\centering
\includegraphics[width=8.5cm]{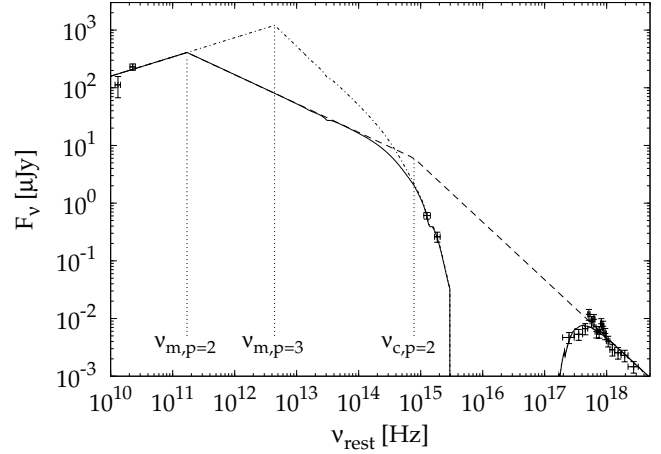}
\caption{Rest-frame spectral energy distribution at $4.1 \times
  10^5$~s.  Two alternative models are shown, depending on whether
  $\nu_{\rm c}$ lies below the optical ($p=2$; solid line) or above
  X-rays ($p=3$; dashed-dotted line).  Characteristic frequencies are
  reported. Both models are extinguished with an LMC2 dust
  profile. The dashed line is the same as the solid line, removed of
  dust extinction.}
\label{f:sedradio}
\end{figure}
%

The expected contribution to the radio flux from the reverse shock
exceeds the observed values by a factor of $10^2$: at the peak time,
$F_{\rm max,r} \approx 80$~mJy.  This is found from
Equation~(\ref{eq:freq}) and from $F_{\rm max,f} \approx F_{\rm
  p,unext} = 1200$~$\mu$Jy, where we used the observed flux density in
the $R$ band at the peak time, $F_{\rm p} = 180 \pm 6$~$\mu$Jy;
removing the intrinsic dust extinction of $A_{V,z} = 0.8$ mag, the
unextinguished value is $F_{\rm p,unext} = 1200$~$\mu$Jy.  From the
deceleration time to the end of the energy injection, it is $F_{\rm
  max,r} \approx F_{\rm max,f}\,\Gamma \propto t^{1-q -(2+q)/8} = t^0$
(Table~\ref{tab:closure_rel} and \citealt{Zhang06}). From that time to
$t_{\rm radio}$, it is $F_{\rm max,r} \propto t^{-1}$
\citep{Kobayashi00,Zhang03}.  It is therefore $F_{\rm max,r}(t_{\rm
  radio}) = 80\,(t_{\rm radio}/t_{\rm b2})^{-1} \approx 20$~mJy.  From
Equation~(\ref{eq:freq}) $\nu_{\rm m,r} \approx \nu_{\rm o}/\Gamma^2 =
1.2 \times 10^{11}$~Hz at the peak time.  Using the temporal scaling
of $\nu_{\rm m,r}$, its value at $t_{\rm radio}$ is found to be
$\nu_{\rm m,r}(t_{\rm radio}) = \nu_{\rm m,r}\,(t_{\rm b2}/t_{\rm
  p})^{-2/3}\,(t_{\rm radio}/t_{\rm b2})^{-3/2} \approx 1$~GHz, where
we used $\nu_{\rm m,r} \propto t^{-2/3}$ for $t < t_{\rm b2}$
(Table~\ref{tab:closure_rel}) and $\nu_{\rm m,r} \propto t^{-3/2}$ for
$t > t_{\rm b2}$ \citep{Kobayashi00}.  This means that around the time
of observations, $\nu_{\rm m,r}$ should have crossed the radio band.
The expected flux should therefore be comparable with $F_{\rm
  max,r}(t_{\rm radio}) \approx 20$~mJy.  Although the passage of the
peak synchrotron (of the reverse shock) through the radio band is
indeed compatible with the radio light curve, the observed flux is
about two orders of magnitude smaller than expected.  Although in
principle self-absorption could explain this, in the following we show
that here it appears unlikely. Another possibility to suppress the
reverse-shock radio emission at late times is assuming a hot flow
(i.e., Poynting-flux dominated).

Equation~(\ref{eq:freq}) is valid as long as we consider energy
injection to the shock by a cold flow (i.e., injection of kinetic
energy). More generally, assuming the same velocity and pressure in
the FS and RS regions, the ratio $F_{\rm max,r}/F_{\rm max,f}$ is
proportional to the corresponding ratio of the number of electrons in
the two shock regions. Considering an extreme case of no new electron
injection in the reverse-shock region, it is $F_{\rm max,r}/F_{\rm
  max,f} \approx N_{\rm e,r}/N_{\rm e,f}\sim r^{-3} \approx
t^{-3\,(2-q)/4}= t^{-1}$.  Similarly, $\nu_{\rm m,r}/\nu_{\rm m,f}
\approx (\gamma_{\rm e,r}/\gamma_{\rm e,f})^2 \approx (\rho_{\rm
  f}/\rho_{\rm r})^2 \approx r^6 \approx t^{3\,(2-q)/2}= t^2$;
$\nu_{\rm c,r} \approx \nu_{\rm c,f} \approx t^{(q-2)/2} \approx
t^{-2/3}$.  These might be more uncertain compared to the cold-flow
results, where we have assumed that the FS and RS regions have
comparable widths.

\item ${\rm max}(\nu_{\rm m},\nu_{\rm c}) < \nu_{\rm o,x}$ ($p=2$).

This requires $\nu_{\rm m,f} \la \nu_{\rm o}$ and $\nu_{\rm c} \la
\nu_{\rm o}$ at $t=t_{\rm p}$.  No energy injection is required to
explain the spectral and temporal properties of the afterglow.  The
characteristic synchrotron frequencies can be expressed by
\citep{Sari98,Kobayashi03a}
\begin{equation}
\nu_{\rm c}(t_{\rm p}) \approx 6.8 \times 10^{12}\, \epsilon_B^{-3/2}
E_{52}^{-1/2} \eta_{0.2}^{1/2} n_0^{-1} \zeta^{-1/2} t_{{\rm
    p},3}^{-1/2}\ \textrm{Hz},
\label{eq:nuc_sari98}
\end{equation}
\begin{equation}
\nu_{\rm m,f}(t_{\rm p}) \approx 1.7 \times 10^{18}\, \epsilon_B^{1/2}
\epsilon_e^{2} E_{52}^{1/2} \eta_{0.2}^{-1/2} \zeta^{1/2} t_{{\rm
    p},3}^{-3/2}\ \textrm{Hz},
\label{eq:numf_sari98}
\end{equation}
%
%
where $\eta_{0.2} = \eta/0.2$, $t_{{\rm p},3} = t_{\rm p}/10^3$~s,
$E_{52} = E_{\rm iso}/10^{52}$~erg, and $\zeta=(1+z)/2.687$.
In particular, it is $\nu_{\rm c} \approx 3.6 \times 10^{12}\,
\epsilon_B^{-3/2} n_0^{-1}$~Hz, $\nu_{\rm m,f} \approx 1.3 \times
10^{18}\,\epsilon_B^{1/2} \epsilon_e^{2}$~Hz, and $\nu_{\rm m,r}
\approx 3 \times 10^{14}\,\epsilon_B^{1/2} \epsilon_e^{2}$~Hz.
%

\begin{figure}
\centering
\includegraphics[width=8.5cm]{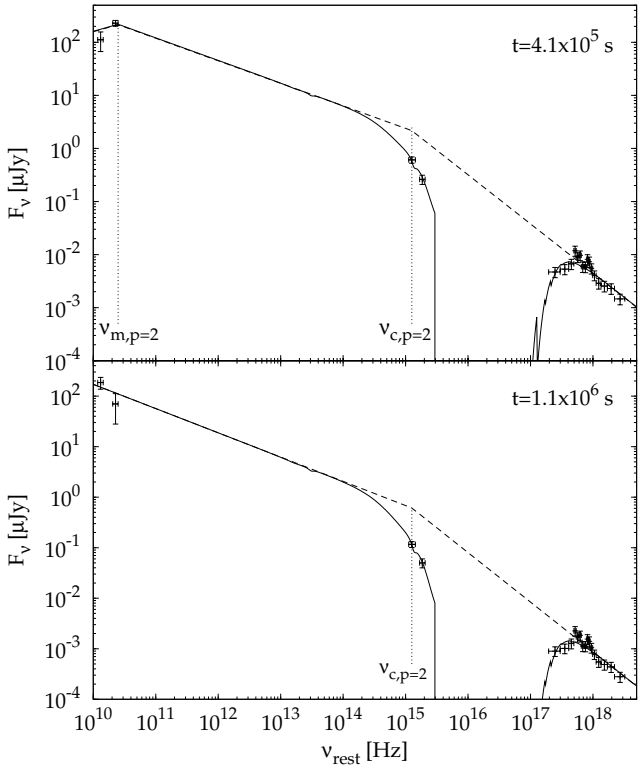}
\caption{{\em Top panel:} rest-frame SED at $4.1 \times 10^5$~s.  The
  solid line shows the synchrotron spectrum with an LMC2
  dust-extinction profile for the $p=2$ case. The corresponding
  injection and cooling frequencies are indicated.  The dashed line
  shows the same unextinguished model.  {\em Bottom panel:} rest-frame
  SED at $1.1 \times 10^6$~s. The data are consistent with being taken
  after the passage of $\nu_{\rm m}$ through the radio band.}
\label{f:sedradio_multi}
\end{figure}
%
The requirements on the characteristic frequencies at the peak time
become $\epsilon_B^{1/2} \epsilon_e^2 \la 4 \times 10^{-4}$ and
$\epsilon_B^{-3/2} n_0^{-1} \la 140$.  A possible solution is
$\epsilon_B \approx 10^{-2}$, $\epsilon_e \approx 3 \times 10^{-2}$,
$n_0 \approx 10$, which gives $\nu_{\rm c} \approx 3.6 \times
10^{14}$~Hz and $\nu_{\rm m,f} \approx 1.2 \times 10^{14}$~Hz.  The
much lower $\nu_{\rm c}$ than the previous case would be mainly due to
a denser environment.

The expected peak flux of the forward shock at the peak frequency
$\nu_{\rm m,f}$ is then $F_{\rm max,f}(t_{\rm p}) \approx (\nu_{\rm
  m,f}/\nu_{\rm c})^{-(p-1)/2}\,(\nu_{\rm c}/\nu_{\rm
  o})^{-p/2}\,F_{\rm p,unext} \approx 3$~mJy.  The contribution of the
reverse shock to the optical peak luminosity in the $R$ band is
comparable: from Equation~(\ref{eq:freq}) the reverse shock peaks at
$\nu_{\rm m,r} \approx \nu_{\rm m,f}/\Gamma^2 \approx 2.4 \times
10^{10}$~Hz with $F_{\rm max,r} \approx \Gamma\,F_{\rm max,f}$ which
scales at $\nu_{\rm m,f}$ by the factor $(\nu_{\rm m,f}/\nu_{\rm
  m,r})^{-(p-1)/2} = \Gamma^{-1}$.  The net result is $F_{r}(\nu_{\rm
  m,f}) \approx F_{f}(\nu_{\rm m,f})$.

At $t=t_{\rm radio}$, the expected luminosity is dominated by the
forward shock. From the frequency and flux-density scalings $\nu_{\rm
  m,f} \propto t^{-3/2}$, $\nu_{\rm c,f} \propto t^{-1/2}$, $F_{\rm
  max,f} \propto t^0$, $\nu_{\rm m,r} \propto t^{-3/2}$, $\nu_{\rm
  c,r} \propto t^{-3/2}$, and $F_{\rm max,r} \propto t^{-1}$
\citep{Zhang03,Mundell07}, we have $\nu_{\rm m,f}(t_{\rm radio})
\approx 3 \times 10^{10}$~Hz, $\nu_{\rm m,r}(t_{\rm radio}) \approx 7
\times 10^{6}$~Hz. This means that the peak of the reverse shock has
already crossed the radio band, while that of the forward shock has
not yet.  The two expected flux densities at 8.46~GHz are $F_f(t_{\rm
  radio}) \approx 3\,(\nu_{\rm radio}/\nu_{\rm m,f}(t_{\rm
  radio}))^{1/3} \approx 2$~mJy and $F_r(t_{\rm radio}) \approx
3\,(\nu_{\rm radio}/\nu_{\rm m,r}(t_{\rm radio}))^{-(p-1)/2}\,(t_{\rm
  radio}/t_{\rm p})^{-1}$~mJy $\approx 0.3$~$\mu$Jy,
respectively. Thus, given that the expected flux from the reverse
shock is 3 orders of magnitude smaller than observed, while that of
the forward shock differs by a factor of few, hereafter we focus on
the latter.

The solid line of Figure~\ref{f:sedradio} shows the result of fitting
the broadband SED at $t_{\rm radio}$ with both $\nu_{\rm m,f}$ and
$\nu_{\rm c,f}$ below the optical and having fixed $p=2$ and the
rest-frame dust extinction. The free parameters are the break
frequencies as well as the normalisation. We found $\nu_{\rm
  m,f}\,(1+z) = 1.7_{-0.8}^{+93} \times 10^{11}$~Hz and $\nu_{\rm
  c,f}\,(1+z) = 8_{-3}^{+5} \times 10^{14}$~Hz ($\chi^2/{\rm dof} =
20.9/17$).  Not only does the value for $\nu_{\rm m,f}$ agree with
expectations, but also the peak flux of $\sim 400$~$\mu$Jy resulting
from the fit is nearly compatible with the value derived from the
radio light curve (Section \ref{sec:disc:radio}).  The downside is the
value for $\nu_{\rm c,f}$: being so close to the optical band, it must
have crossed it at some time $t \la t_{\rm radio}$. This would imply
$\beta_{\rm o} = (p-1)/2 = 0.5$ for $t < t_{\rm radio}$, which is
clearly not true. A possible solution could be to assume a cooling
frequency that increases with time, but given that we ruled out a wind
environment, this option is not acceptable either. Another problem
concerns the crossing time of $\nu_{\rm c,f}$ through the radio band,
observed immediately afterward at $t_{\rm radio,p} \approx 1.1\,t_{\rm
  radio}$ (Section \ref{sec:disc:radio}).

We tried to decrease the latter problem by fixing $\nu_{\rm m,f}$ to
the value expected from the time it crossed the radio band --- that
is, by imposing $\nu_{\rm m,f} = \nu_{\rm radio}\,(t_{\rm
  radio}/t_{\rm radio,p})^{-3/2} = 9.4$~GHz. We allowed all the
remaining parameters to vary. The result is shown in the top panel of
Figure~\ref{f:sedradio_multi}.  The best-fitting parameters are
$\nu_{\rm c,f}\,(1+z) = (1.3 \pm 0.1) \times 10^{15}$~Hz, $\beta_{\rm
  ox} = 0.92 \pm 0.04$ and $A_{V,z} = 0.56 \pm 0.10$ mag. Although the
spectral index is slightly harder than that found at previous epochs,
both the normalisation and the slope of the radio-to-optical spectrum
fit in the expected broadband modelling well. This explains both the
spectrum and light curve of the radio observations.  The bottom panel
of Figure~\ref{f:sedradio_multi} shows the SED at $t = 1.1 \times
10^6$~s, when $\nu_{\rm m,f}$ has just crossed the radio band. Apart
from the same issue with $\nu_{\rm c,f}$ already mentioned, this is in
remarkable agreement with expectations: $\nu_{\rm c,f}\,(1+z) = (1.3
\pm 0.1) \times 10^{15}$~Hz, $\beta_{\rm ox} = 0.98 \pm 0.05$, and
$A_{V,z} = 0.72 \pm 0.10$ mag.  Yet, the inferred temporal evolution
of the cooling frequency remains an issue. In this case, this would
lie within the optical bands and the aforementioned argument still
applies.

Overall, the $p=2$ case works better than $p=3$ and can account for
more observed properties of the broadband afterglow evolution. Still,
the derived evolution of the cooling frequency with time conflicts
with a homogeneous environment, the only one compatible with the data.

\subsubsection{Self-absorption}
\label{sec:disc:selfabs}

So far we assumed a negligible effect due to self-absorption in the
observed radio flux.  Should the radio flux be self-absorbed, from
Figure~\ref{f:sedradio} both cases could be compatible with having a
high value of $F_{\rm max}$. In particular, for the $p=2$ case the
temporal evolution of $\nu_{\rm c,f}$ would no longer be an issue,
because it could lie well below $\nu_{\rm o}$ at $t = t_{\rm radio}$.
To significantly suppress the radio flux, one should require $\nu_{\rm
  c} \ll \nu_{\rm o}$ already at the peak time; however, this would
imply a much larger optical luminosity and an unusually high energy
budget.  This seems very unlikely, given that the optical luminosity
of GRB~080603A, corrected for the dust extinction, already lies in the
mid-to-bright end of the observed optical afterglow distribution (Section
\ref{sec:disc:dust} and Figure~\ref{f:kann_z1}).

Although at lower frequencies and early times self-absorption can
significantly suppress the flux, this does not explain these
observations unless one makes extreme assumptions.  A simple estimate
of the maximum flux is that of a black body with the forward-shock
temperature \citep{Sari99,Kobayashi00a,Mundell07}, which at the peak
time in the optical ($t = t_{\rm p}$) is given by
\begin{eqnarray}
F_{\nu,{\rm BB}}(t_{\rm p}) & \approx & \pi (1+z) \nu^2
\epsilon_e m_p \Gamma^2\,\Big(\frac{R_\bot}{D_L}\Big)^2\\ & \approx
& 500\,\Big(\frac{\nu}{\nu_{\rm radio}}\Big)^2
\epsilon_{e,-2} n_0^{-1/2}\ \ \textrm{$\mu$Jy},
\label{eq:F_BB}
\end{eqnarray}
where $R_\bot \approx 4.6 \Gamma c t_{\rm p}$ is the observed fireball
size and the dependence on $n_0$ is inherited from
Equation~(\ref{eq:gamma_ism}).  The value derived from
Equation~(\ref{eq:F_BB}) initially increases as $\sim t^{1/2}$, and
steepens to $\sim t^{5/4}$ after $\nu_{\rm m,f}$ crosses the observed
frequency; thus, at the time of radio observations the black-body
flux-density limit expressed by Equation~(\ref{eq:F_BB}) increases by
a factor of $\sim 20$. This can hardly explain the observed radio flux
unless one assumes $\epsilon_e \approx 10^{-4}$ and/or a high-density
environment ($n \approx 400$~cm$^{-3}$).  This value for $\epsilon_e$
would imply that $\nu_{\rm m}$ is much below the optical bands
($\nu_{\rm m,f} \approx 10^{8\div9}$~Hz) at the peak time and,
consequently, an unreasonably high value for $F_{\rm max,f}$.
\end{itemize}

\subsubsection{Off-axis jet}
\label{sec:offaxis_jet}

In the off-axis jet interpretation, the afterglow rise and peak do not
mark the fireball deceleration, but are the result of a geometric
effect due to an observer angle, $\theta_{\rm obs}$, being larger than
the jet opening angle, $\theta_{\rm j}$: as the jet decelerates, the
relativistic beaming angle $1/\Gamma$ progressively increases,
resulting in a rising light curve as seen from the off-axis observer,
provided that the jet structure is such that the energy and $\Gamma$
drop sharply at $\theta > \theta_{\rm j}$.  The peak is reached when
$\Gamma\,(\theta_{\rm obs} - \theta_{\rm j}) \approx 1$, as the
sightline enters the beaming cone at the edge of the jet; finally, the
decay asymptotically approaches the light curve for an on-axis
observer \citep{Granot05}.

We can estimate $\theta_{\rm obs}$ from the peak time as follows:
$\theta_{\rm obs}-\theta_{\rm j} \approx 1/\Gamma(t_{\rm
  p})$. Interpreting the late-time break as being due to the jet, it
is $\theta_{\rm j} \approx 1/\Gamma(t_{\rm j})$, from which we
estimated $\theta_{\rm j} \approx 5\fdg7$ (Section
\ref{sec:disc:dust}). Assuming that the deceleration of the jet occurs
before the peak, we can assume the temporal scaling of the bulk
Lorentz factor as $\Gamma \propto t^{-3/8}$ for an adiabatic cooling
\citep{Rees92}:
\begin{equation}
\frac{\theta_{\rm obs} - \theta_{\rm j}}{\theta_{\rm j}}\ \approx
\frac{\Gamma(t_{\rm j})}{\Gamma(t_{\rm p})}\ \approx \Big(\frac{t_{\rm
    j}}{t_{\rm p}}\Big)^{-3/8}\ \approx 0.2,
\label{eq:theta_obs}
\end{equation}
so $(\theta_{\rm obs}-\theta_{\rm j}) \approx 1^\circ$.

Given the steep rise observed for GRB~080603A, we conveniently
consider the case of a sharp-edged, homogeneous jet seen at
$\theta_{\rm obs} > \theta_{\rm j}$. The observed peak energy of the
prompt emission, $E_{\rm p}$, falls off as $b^2$, where $b =
\Gamma\,(\theta_{\rm obs} - \theta_{\rm j})$, while the observed
energy falls off rapidly as $b^6$ \citep{Granot02}.  From
Equation~(\ref{eq:theta_obs}), $b \approx 2\,\Gamma_{100}$, implying
the following on-axis values: $E_{\rm p}(\theta_{\rm obs} = 0) \approx
600\,\Gamma_{100}^2$~keV and, and $E_{\rm iso}(\theta_{\rm obs} = 0)
\approx 10^{54}\,\Gamma_{100}^6$~erg; here, $\Gamma$ is the initial
value of the bulk Lorentz factor.  These estimates are only
illustrative and show that, in principle, this scenario could work for
reasonable values of the physical parameters in the case of a uniform
sharp-edged jet.

However, more realistically the jet cannot be exactly uniform with
very sharp edges; in particular, $\Gamma$ is expected to be lower at
the edge and higher at the jet core.  Hydrodynamic simulations have
shown that, particularly for $1 \la \theta_{\rm obs}/\theta_{\rm j}
\la 2$, very shallow rises or decays of the early-afterglow light
curves are expected for realistic jet structure and dynamics
\citep{Granot02,Eichler06}.

Alternatively, structured jets \citep{Zhang02} with most of the energy
concentrated in the core can reproduce steep rises in the light curves
\citep{Kumar03,Granot05,Eichler06}.

Following \citet{Panaitescu08}, assuming an angular profile for energy
as $\mathcal{E}(\theta) \propto (\theta/\theta_{\rm j})^{-q}$, a $\sim
t^3$ rise would be obtained for $q=4$ and $\theta_{\rm
  obs}/\theta_{\rm j} \approx 1.5\div2.5$.  GRB~080603A lies within
the 1$\sigma$ region of the correlation between peak time and peak
flux for fast-rising afterglows, also discussed by
\citet{Panaitescu11} (Section \ref{sec:disc:ag_onset}; see Fig.~2 of
\citealt{Panaitescu08}).

The scenario of a viewing angle slightly larger than the jet angle
seems to be consistent with the nature of a typical GRB, in contrast
to the classes of spectrally softer events, such as the so-called
X-ray rich GRBs (XRRs) or X-ray flashes (XRFs; e.g.,
\citealt{Sakamoto08} and references therein).  Larger viewing angles
are expected to be associated with softer observed events
\citep{Yamazaki02,Granot02,Granot05}, and the off-axis interpretation
for events of this type appears to be favoured (e.g.,
\citealt{Guidorzi09}).  In this context, GRB~080603A could represent a
soft/intermediate classical GRB with a typical jet opening angle and
viewed with a comparable viewing angle.

\subsection{Prompt optical/$\gamma$-ray emission: inverse Compton?}
\label{sec:disc:flash}

Before the onset of the afterglow marked by the steep rise at the end
of the $\gamma$-ray prompt emission, the optical flux detected
simultaneously with the second and last $\gamma$-ray pulse is unlikely
to be synchrotron radiation of the shocked ISM.  As noted in Section
\ref{sec:flash}, the spectral index measured during the prompt
emission in the $\gamma$-ray band is likely to be an intermediate
value between the typical low-energy and high-energy photon indices of
a Band function. In this respect, the optical emission could be
consistent with the extrapolation of the $\gamma$-ray spectrum down to
the optical band.  A cross-correlation study between the optical and
$\gamma$-ray profiles would certainly settle this issue; however, in
practice this is not possible because of the coarse optical coverage,
which gave only three points separated by gaps in between
(Fig.~\ref{f:lc_gamma}).

The variety of observed behaviours in other GRBs is rich: the prompt
optical was observed to be uncorrelated with the ongoing high-energy
emission for GRB~990123 \cite{Akerlof99} (e.g., see also GRB~060111B,
\citealt{Klotz06,Stratta09}; GRB~080607, \citealt{Perley11}), whereas
a strong correlation was observed, for example, for GRB~050820A
\citep{Vestrand06}, superposed on the onset of the afterglow
\citep{Cenko06}. Similar cases of some degree of correlation between
the $\gamma$-ray prompt and optical emissions are GRB~041219A
\citep{Vestrand05,Blake05}, GRB~060526 \citep{Thoene10}, and
GRB~080319B \citep{Racusin08,Beskin10}.  One of the most common cases
is that the prompt optical observations, typically starting during the
final part of the $\gamma$-ray emission, suggest the transition from
the inner-engine activity to the multi-band afterglow onset; see, for
instance, GRB~051111 \citep{Yost07}, GRB~081008 \citep{Yuan10},
GRB~081126 \citep{Klotz09}, and GRB~080928 \citep{Rossi11}. In one
case, a strong optical flare incompatible with an external-shock
origin was observed before the afterglow onset \citep{Greiner09}.  In
some other cases, the optical profile is dominated by the onset of the
external shock of the ejecta through the ISM (e.g., GRB~080810,
\citealt{Page09}; GRB~061007, \citealt{Rykoff09,Mundell07}).

Evidence has also been reported for a sizable temporal lag of a few
seconds between the optical and high-energy profiles
\citep{Klotz09,Rossi11,Beskin10}. This potentially represents a strong
clue to explain the prompt-emission mechanism.

Comparison of the optical and $\gamma$-ray fluxes is also useful for
establishing the possible link: while some GRBs have an
optical-to-$\gamma$ spectral index $\beta_{{\rm opt}-\gamma}$
compatible with $\beta_{\gamma}$, as can be the case for GRB~080603A,
other events show an excess of optical emission with respect to the
extrapolation of the high-energy spectrum: up to $10^4$ times larger,
as in the case of GRB~080319B \citep{Racusin08}. By contrast, there
are also cases in which the optical flux lies below the high-energy
spectrum extrapolation, such as GRB~050401 \citep{Rykoff05,Yost07}.
The latter situation does not necessarily imply different origins or
mechanisms for optical and high-energy emissions,
but could merely be due to either dust extinction (as is at
least partially the case for GRB~050401, \citealt{Kann10})
or a synchrotron spectrum peaking between the two energy ranges.

GRB~080603A is an example where both components (internal activity and
afterglow onset) are clearly temporally separated.  Although no firm
conclusion can be drawn on the possible existence of a temporal lag
between optical and $\gamma$-ray photons, this GRB resembles
GRB~081126 \citep{Klotz09}: both $\gamma$-ray profiles consist of two
disjoint FRED-like pulses, the last of which is observed
simultaneously with an optical flash, followed by the afterglow onset.

Following \citet{Piran09}, we tested whether the $\gamma$-ray prompt
emission of GRB~080603A can be explained in terms of an IC process by
a population of relativistic electrons on low-energy seed photons. The
same electrons would also upscatter the $\gamma$-ray photons to
GeV--TeV energies. To avoid the energy crisis (i.e., when most of the
energy is released in the GeV--TeV range owing to a large $Y$
parameter), we used the simultaneous optical and $\gamma$-ray flux
densities to constrain the bulk Lorentz factor $\Gamma$.  Let
$\gamma_{\rm e}$ be the Lorentz factor of electrons within the fluid
rest frame, $\nu_{\rm L}$ the (GRB rest-frame) peak frequency of the
lower spectral component (i.e., of the seed $\nu F_\nu$ spectrum), and
$F_{\rm L}$ the corresponding peak flux. Also, let $\nu_{\rm opt} =
\nu_{\rm o}\,(1+z) = 1.34 \times 10^{15}$~Hz be the rest-frame
frequency corresponding to the observed optical band. $F_{\rm L}$ can
then be expressed as
\begin{equation}
F_{\rm L}\ =\ (\nu_{\rm L}/\nu_{\rm opt})^{-\beta}\,F_{\rm opt},
\label{eq:F_L}
\end{equation}
where $F_{\rm opt} = 160$~$\mu$Jy is the unextinguished flux of the
optical flash.  Two possible cases are considered: for the UV (IR)
solution, corresponding to $\nu_{\rm L} > \nu_{\rm opt}$ ($\nu_{\rm L}
< \nu_{\rm opt}$), we assume three possible values for the spectral
index: $\beta=0$, $-0.5$, and $-1$ ($\beta=1$, 1.5, and 2). The
Compton parameter $Y_{\rm L}$ in the first IC scattering is
\begin{equation}
Y_{\rm L}\ =\ \Big(\frac{\nu_\gamma\,F_\gamma}{\nu_{\rm opt}\,F_{\rm
    opt}}\Big)\, \Big(\frac{\nu_{\rm L}}{\nu_{\rm
    opt}}\Big)^{-(1-\beta)}\ \approx 10^4\ \Big(\frac{\nu_{\rm
    L}}{\nu_{\rm opt}}\Big)^{-(1-\beta)},
\label{eq:Y_L}
\end{equation}
where $h\nu_\gamma = 84\,(1+z) = 226$~keV and $F_\gamma = 41$~$\mu$Jy
(Table~\ref{tab:N05}).  The first-order IC scattering is not in the
Klein-Nishina (KN) regime, so it is $\nu_{\gamma}/\nu_{\rm L} =
\gamma_{\rm e}^2$, or, equivalently,
\begin{equation}
\gamma_e\ = \ \Big(\frac{\nu_\gamma}{\nu_{\rm
    opt}}\Big)^{1/2}\ \Big(\frac{\nu_{\rm opt}}{\nu_{\rm
    L}}\Big)^{1/2} =\ 200\ \Big(\frac{\nu_{\rm opt}}{\nu_{\rm
    L}}\Big)^{1/2}\quad .
\label{eq:gamma_e}
\end{equation}
The second-order IC scattering might be in the KN regime, so 
$Y_{\rm H}$ is
\begin{equation}
Y_{\rm H}\ =\ \Big(\frac{\nu_\gamma\,F_\gamma}{\nu_{\rm opt}\,F_{\rm
    opt}}\Big)\, \Big(\frac{\nu_{\rm L}}{\nu_{\rm
    opt}}\Big)^{-(1-\beta)}\ {\rm min}(1, \xi^{-2}),
\label{eq:Y_H}
\end{equation}
where $\xi = \,(\gamma_{\rm e}/\Gamma)\,h\nu_\gamma/m_{\rm e}\,c^2$ is
the correction factor in the KN regime. The energy of the upscattered
photons is
\begin{equation}
h\nu_{\rm H}\ =\ 9\ \textrm{GeV}\ \Big(\frac{\gamma_{\rm
    e}}{200}\Big)^2\ {\rm min}\Big(1, \frac{\Gamma\,m_{\rm
    e}c^2}{\gamma_{\rm e}\,h\nu_\gamma}\Big),
\label{eq:nu_H}
\end{equation}
and the corresponding Compton parameter $Y_{\rm H}$ is
\begin{equation}
Y_{\rm H}\ =\ 10^4\ \Big(\frac{\nu_{\rm L}}{\nu_{\rm
    opt}}\Big)^{-(1-\beta)}\ {\rm min}\Big[1,
  \Big(\frac{\Gamma\,m_{\rm e}c^2}{\gamma_{\rm
      e}\,h\nu_\gamma}\Big)^2\Big].
\label{eq:Y_H2}
\end{equation}
%
%
\begin{figure}
\centering
\includegraphics[width=8.5cm]{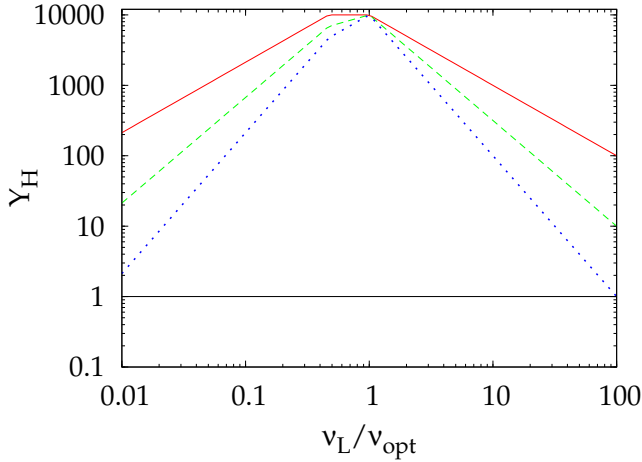}
\caption{$Y_{\rm H}$ as a function of $\nu_{\rm L}/\nu_{\rm opt}$ for
  $\Gamma = 130$ and (from top to bottom) $\beta=0$, $-0.5$, and $-1$
  for $\nu_{\rm L} > \nu_{\rm opt}$, and (from top to bottom)
  $\beta=1$, 1.5, 2 for $\nu_{\rm L} < \nu_{\rm opt}$ (adapted from
  \citealt{Piran09}).}
\label{f:IC1}
\end{figure}
%

The energy-crisis problem mainly resides in the large value for
$Y_{\rm H}$.  Figure~\ref{f:IC1} shows $Y_{\rm H}$ as a function of
$\nu_{\rm L}$ for the values of the spectral index considered above
and assuming for $\Gamma$ the value we derived in
Equation~(\ref{eq:gamma_ism}).  There are two solutions for which the
released energy is not an issue: the UV solution requires $\beta \le
-1$ and $\nu_{\rm L} > 30\,\nu_{\rm opt}$, equivalent to $\gamma_{\rm
  e} < 37$ (Eq.~(\ref{eq:gamma_e})). This solution is characterised by
a negligible KN suppression. The total energy is at least
$3\,E_\gamma$, given that $Y_{\rm H} \approx Y_{\rm L}$. Other
problems with the UV solutions are (i) a low efficiency due to the low
$\gamma_{\rm e}$, because protons would carry at least a factor of
$m_{\rm p}/\gamma_{\rm e}\,m_{\rm e}$ more energy than electrons,
unless one requires pair loading; and (ii) at such low values of
$\gamma_{\rm e}$, $\nu_{\rm L}$ increases and its flux $F_{\rm L}$ is
limited not only by $F_{\rm opt}$, but also by the prompt soft X-ray
observations. Whenever available, these data rule out the UV solution
\citep{Piran09}.  Although for GRB~080603A there are no prompt X-ray
observations, the UV solution appears contrived.

The IR solution requires $\beta \ge 2$ and $\nu_{\rm L} \la
0.01\,\nu_{\rm opt}$, equivalent to $\gamma_{\rm e} \ga 2000$. This
solution is characterised by a high value for $Y_{\rm L}$, and a
relatively small value for $Y_{\rm H}$, because of the KN suppression.
This becomes important for $\nu_{\rm L} < 0.46\,\nu_{\rm opt}$, as shown
by the break in Figure~\ref{f:IC1}.
%
\begin{figure}
\centering
\includegraphics[width=8.5cm]{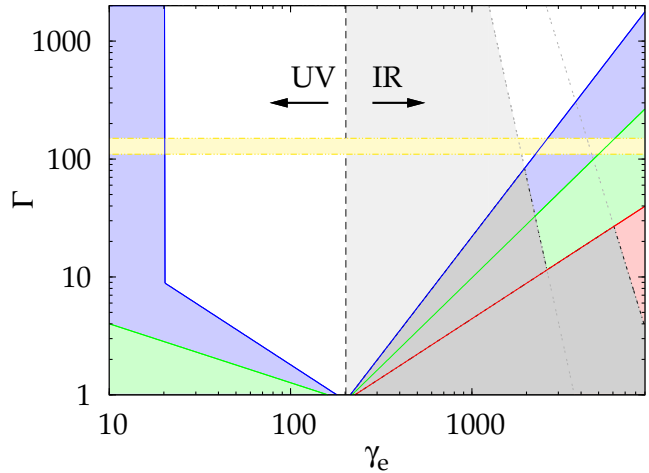}
\caption{The allowed (shaded) phase space characterised by $Y_{\rm H}
  \le 1$ (from bottom to top) for $\beta=0$, $-0.5$, and $-1$ for
  $\nu_{\rm L} > \nu_{\rm opt}$ and (from bottom to top) $\beta=1$,
  1.5, 2 for $\nu_{\rm L}<\nu_{\rm opt}$. The two decreasing functions
  in the IR solution mark the self-absorption limits for $\beta=2$
  (grey area; lower $\gamma_{\rm e}$ range) and $\beta=1$ (higher
  $\gamma_{\rm e}$ range), respectively.  The two darkest areas
  partially overlapping each other are the intersection of the allowed
  regions for $\beta=2$ and for $\beta=1$, respectively.  The interval
  $\Gamma = 130 \pm 20$ (estimated from the peak in the afterglow
  light curve) is highlighted.}
\label{f:IC2}
\end{figure}
%
On the other hand, for very low $\nu_{\rm L}/\nu_{\rm opt}$ the
expected $F_{\rm L}$ also increases for a given $F_{\rm opt}$ and
self-absorption can represent an issue. To avoid this, one has to
require $F_{\rm sa}(\nu_{\rm L}) > F_{\rm L}$, where $F_{\rm
  sa}(\nu_{\rm L})$ is the black-body flux for a local temperature
$kT \approx \Gamma\,\gamma_{\rm e}\,m_{\rm e}c^2$,
\begin{equation}
F_{\rm sa}(\nu_{\rm L})\ =\ \frac{2 \nu_{\rm L}^2}{c^2} \gamma_{\rm
  e} m_{\rm e} c^2\, \frac{R^2}{4 \Gamma d_{\rm L}^2},
\label{eq:F_sa}
\end{equation}
where $R$ is the radius of the source and $d_{\rm L} = 12.74$~Gpc is the
luminosity distance of GRB~080603A. From this requirement on $F_{\rm
  L}$ we can constrain $\Gamma$ for $\nu_{\rm L} < \nu_{\rm opt}$:
\begin{equation}
\Gamma < \frac{1}{2}\,\Big(\frac{R}{d_{\rm
    L}}\Big)^2\,\nu_\gamma^{2+\beta}\,\frac{\nu_{\rm
    opt}^{-\beta}}{F_{\rm opt}} \,m_{\rm e} \gamma_{\rm
  e}^{-(2\,\beta+3)}\quad .
\label{eq:G_sa}
\end{equation}
We conservatively assume $R = 10^{17}$~cm.  The corresponding allowed
$\Gamma$--$\gamma_{\rm e}$ phase space is shown in Figure~\ref{f:IC2}
in the IR-solution domain, enclosed by the decreasing curves on the
right-hand side.  Combining this with the $Y_{\rm H} \le 1$ regions
and with the corresponding values for the spectral index shows that
the value measured for $\Gamma$ in Equation~(\ref{eq:gamma_ism}) does
not overlap with any total allowed region (darkest areas).

We conclude that the prompt optical and $\gamma$-ray data are not
compatible with an IC origin for the latter as a result of
upscattering of seed NIR/UV photons causing the prompt optical flash.

\subsection{Dust extinction, luminosity, and energetics}
\label{sec:disc:dust}

\begin{figure}
\centering
\includegraphics[width=8.5cm]{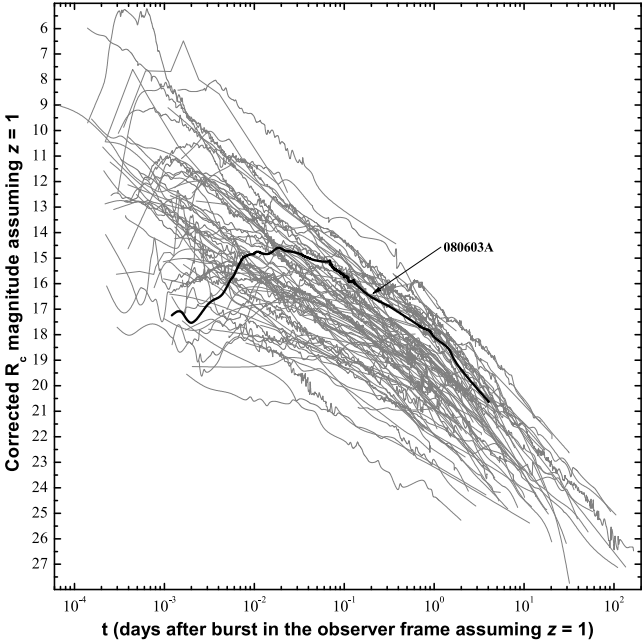}
\caption{The optical afterglow light curve of GRB~080603A (thick line)
  moved to a common redshift of 1 and compared with the analogous
  sample of long GRBs (grey lines) taken from \citet{Kann10}. All
  afterglow curves have been corrected for dust extinction.}
\label{f:kann_z1}
\end{figure}

Figure~\ref{f:kann_z1} shows the optical afterglow curve of
GRB~080603A moved to a common $z=1$ and corrected for the large dust
extinction due to the sightline within the host galaxy as described by
\citet{Kann06}. The sample of other GRBs shown is taken from
\citet{Kann10}.  The afterglow of GRB~080603A ranks among the
mid/bright GRBs.  Although not a dark GRB according to the
$\beta_{\rm ox} < 0.5$ definition, this is a fair example of an optically
observed dim burst mainly because of the large amount of dust within
the host galaxy: the value $A_{V,z} = 0.8$ mag is indeed, among those
measured with good accuracy, one of the largest observed so far
\citep{Kann10}. This agrees with the findings from GRB host-galaxy
studies \citep{Perley09}, samples of GRBs with multi-colour
photometric datasets \citep{Cenko09}, and some individual GRBs
\citep{Perley11}.  Furthermore, the SED we built is one of the very
few which clearly favours an LMC2 extinction profile, with possible
evidence for the presence of the 2175\,\AA\ bump that has rarely been observed
in GRB afterglows \citep{Kruhler08,Eliasdottir09,Perley11}.
%
%

We could not directly measure the peak energy $E_{\rm p}$ of the
prompt $\gamma$-ray spectrum; nevertheless, from the intermediate
value of the photon index, $\Gamma = 1.6$, we can conservatively
assume $E_{\rm p,i} = 160_{-130}^{+920}$~keV. Combining this with the
isotropic-equivalent radiated energy, $E_{\rm iso} = (2.2 \pm 0.8)
\times 10^{52}$~erg, GRB~080603A does not violate the $E_{\rm
  p,i}$--$E_{\rm iso}$ \citep{Amati02,Amati10} relation.

Interpreting the late-time break as being due to a jet, we can provide
an estimate of its opening angle, which in the ISM case turns out to
be $\theta_{\rm j} = 5\fdg7 (-1\fdg2, +1\fdg5)$ \citep{Sari99b}.  We
assumed standard values for the energy conversion efficiency,
$\eta_\gamma = 0.2$ and for the particle density of the circumburst
environment, $n = 3$~cm$^{-3}$. The collimation-corrected released
energy is $E_{\gamma} = (1.1 \pm 0.4) \times 10^{50}$~erg. This agrees
well with the expectations of the $E_{\rm p,i}$--$E_{\gamma}$ relation
\citep{Ghirlanda04,Ghirlanda07}, although the large uncertainty in
$E_{\rm p,i}$ leaves this open. Calculating the analogous values in
the case of a wind profile \citep{Nava06}, these are very similar,
although this density profile is disfavoured from the afterglow rise
slope (Section \ref{sec:disc:ag_onset}): $\theta_{\rm j} = 5\fdg1 (-0\fdg9,
+1\fdg4)$ and $E_{\gamma} = (0.9 \pm 0.3) \times 10^{50}$~erg for the
wind.  Such an opening angle for the possible jet is very typical
\citep{Zeh06,Nava06,Racusin09}.

\section{Summary and conclusions}
\label{sec:conc}

GRB~080603A exhibits a number of properties which allow us to strongly
test many aspects of the prompt and afterglow emission standard
model. Our broadband dataset spans from the prompt $\gamma$-rays out
to the radio band 13 days post burst, and also includes spectroscopic
observations of the afterglow as well as late-time photometry of the
host galaxy.  The main features of GRB~080603A are as follows:
\begin{itemize}
\item a faint ($R \approx 20$~mag) optical flash coincident with the
  last episode of a two-pulse, 150-s long $\gamma$-ray prompt burst;
\item a subsequent achromatic steep rise and peak around 1600~s, which
  probably marks the afterglow onset;
\item no evidence for reverse-shock emission;
\item ISM circumburst environment favoured from afterglow modelling;
\item peak in the radio light curve detected at $\sim$5~days, likely
  caused by the passage of the synchrotron spectrum peak;
\item late-time break in the afterglow light curve, interpreted as a
  jet break; the corresponding opening angle is $\theta_{\rm j} =
  5\fdg7$~$(-1\fdg2, +1\fdg5)$;
\item isotropic-equivalent $\gamma$-ray released energy $E_{\rm iso} =
  (2.2 \pm 0.8) \times 10^{52}$~erg and a collimation-corrected value
  of $E_{\gamma} = (1.1 \pm 0.4) \times 10^{50}$~erg, both typical for
  long GRBs;
\item remarkable dust extinction within the host galaxy, $A_{V,z} =
  0.80 \pm 0.13$ mag, that can be fit with an LMC2 profile (with marginal
evidence for the 2175\,\AA\ bump), and cannot be fit by the average MW and SMC
  curves, at variance with most GRB extinction profiles;
\item a comparable host-galaxy extinction (LMC; $A_V = 0.77$ mag) is
  required for fitting the host SED, possibly suggesting that the
  afterglow is being extinguished by a typical sightline through the
  host ISM;
\item extinction-corrected optical afterglow luminosity that lies in
  the mid-to-bright end of the distribution of GRBs at known redshift;
  and
\item projected offset from the host galaxy centre $<6$~kpc, well
  within the offset distribution of long GRBs.
\end{itemize}

Overall, the standard afterglow model seems to account for almost all
of the observed properties of the broadband afterglow evolution. In
particular, the best solution is given by an electron energy index
distribution of $p=2$, with both cooling and injection frequencies
below the optical band at the time of the peak. However, the temporal
evolution of the characteristic frequencies of the synchrotron
spectrum can hardly be explained assuming typical values for the
microphysical parameters.

We have constrained and crosschecked the Lorentz factor in different
ways: interpreting the optical afterglow peak as the fireball
deceleration yields $\Gamma = 130 \pm 20$.  Secondly, following
\citet{Zou10}, we exploited the presence of a quiescent time between
the two $\gamma$-ray pulses and derived an upper limit to $\Gamma$ of
220.  Finally, we focused on the optical and $\gamma$-ray prompt
radiation to test whether inverse Compton could be a viable mechanism
to explain the GRB, and found that the allowed range for $\Gamma$ is
not compatible with the estimate derived from the afterglow properties
(Fig.~\ref{f:IC2}; \citealt{Piran09}). Alternative interpretations of
the optical flash, such as a reverse-shock origin, are excluded by the
gap between the flash and the afterglow onset. Instead, an interesting
possibility is that of an optical flash due to internal shocks with a
narrow distribution of ejecta Lorentz factors at larger radii. This
would explain both the optical flash being temporally disjoint from
the afterglow onset and the peaky profile of the latter
\citep{Meszaros99,Panaitescu11}.

As an alternative to interpreting the optical afterglow peak as the
fireball deceleration, the off-axis scenario requires an observer line
of sight slightly off the jet cone, with $(\theta_{\rm
  obs}-\theta_{\rm j}) \approx 1^\circ$. The observed steep rise
requires a structured jet with most of the energy in the jet core,
with $q \approx 4$ for an angular profile modelled as
$\mathcal{E}(\theta) \propto (\theta/\theta_{\rm j})^{-q}$
\citep{Panaitescu08}.

Summing up, the clues derived for GRB~080603A to understand the nature
of the prompt emission of GRBs highlight the importance of combining
broadband campaigns and deep ($R > 20$~mag), rapid follow-up
observations capable of exploring the faint end of the prompt optical
emission.

\section*{Acknowledgments}

C.G. acknowledges ASI for financial support (ASI-INAF contract
I/088/06/0). D.A.K. acknowledges support by grant DFG Kl766/16-1,
U. Laux for obtaining the TLS observations, and S. Ertel for kindly
granting him observing time.  J.X.P. is partially supported by
NASA/{\it Swift} grants NNG08A099G, NNX10AI18G and US National Science
Foundation (NSF) CAREER grant AST--0548180.  A.G. acknowledges
founding from the Slovenian Research Agency and from the Centre of
Excellence for Space Sciences and Technologies SPACE-SI, an operation
partly financed by the European Union, the European Regional
Development Fund, and the Republic of Slovenia, Ministry of Higher
Education, Science, and Technology.  A.V.F.'s group at U.C. Berkeley
has been supported by NASA/{\it Swift} Guest Investigator grants
NNX09AL08G, NNX10AI21G, and GO-7100028, as well as by NSF grant
AST-0908886 and the TABASGO Foundation.  KAIT and its ongoing
operation were made possible by donations from Sun Microsystems, Inc.,
the Hewlett-Packard Company, AutoScope Corporation, Lick Observatory,
the NSF, the University of California, the Sylvia \& Jim Katzman
Foundation, and the TABASGO Foundation. Some of the data presented
herein were obtained at the W.M. Keck Observatory, which is operated
as a scientific partnership among the California Institute of
Technology, the University of California, and NASA; the observatory
was made possible by the generous financial support of the W.M. Keck
Foundation.


\begin{table*}
\centering
  \caption{Photometric data set of the NIR/optical afterglow of 
GRB~080603A. Uncertainties are 1$\sigma$.}
  \label{tab:phot}
  \begin{tabular}{rcrclcrcrclc}
\hline
Time$^{\mathrm{a}}$ & Telescope & Exp. & Filter &
Magnitude$^{\mathrm{b}}$ & Flux$^{\mathrm{c}}$ & Time$^{\mathrm{a}}$ &
Telescope & Exp. & Filter & Magnitude$^{\mathrm{b}}$ &
Flux$^{\mathrm{c}}$\\ (s) & & (s) & & & ($\mu$Jy) & (s) & & (s) & & &
($\mu$Jy)\\
\hline
    105        &  FTN      &    10    &    $R$    &  $20.78\pm0.26$ & $ 16.7\pm 4.5$ & 94762        &  FTN      &   600    &   $i'$    &  $21.33\pm0.12$  & $ 11.6\pm 1.4$\\
    136        &  FTN      &    10    &    $R$    &  $20.45\pm0.25$ & $ 22.7\pm 5.9$ &   286        &  KAIT     &    15    &    $I$    &  $> 17.00$       & $< 627$\\
    167        &  FTN      &    10    &    $R$    &  $21.40\pm0.52$ & $  9.4\pm 5.8$ &   414        &  KAIT     &    45    &    $I$    &  $> 17.30$       & $< 476$\\
    558        &  FTN      &    30    &    $R$    &  $18.93\pm0.05$ & $ 91.8\pm 4.3$ &   604        &  KAIT     &    60    &    $I$    &  $17.90\pm0.50$  & $274\pm160$\\
    883        &  FTN      &    60    &    $R$    &  $18.31\pm0.03$ & $162.6\pm 4.6$ &   860        &  KAIT     &   180    &    $I$    &  $17.60\pm0.17$  & $361\pm61$\\
   1381        &  FTN      &   120    &    $R$    &  $18.30\pm0.03$ & $164.1\pm 4.6$ &  1165        &  KAIT     &    60    &    $I$    &  $> 17.50$       & $< 396$\\
   2112        &  FTN      &   180    &    $R$    &  $18.27\pm0.03$ & $168.7\pm 4.7$ &   252        &  FTN      &    10    &    $B$    &  $21.73\pm0.36$  & $ 10.3\pm 4.1$\\
   2927        &  FTN      &   120    &    $R$    &  $18.40\pm0.03$ & $149.6\pm 4.2$ &   478        &  FTN      &    30    &    $B$    &  $20.66\pm0.10$  & $ 27.6\pm 2.7$\\
   3745        &  FTN      &   180    &    $R$    &  $18.61\pm0.03$ & $123.3\pm 3.5$ &   772        &  FTN      &    60    &    $B$    &  $19.73\pm0.04$  & $ 65.0\pm 2.4$\\
   5846        &  FTN      &    72    &    $R$    &  $18.72\pm0.05$ & $111.4\pm 5.3$ &  1209        &  FTN      &   120    &    $B$    &  $19.76\pm0.03$  & $ 63.2\pm 1.8$\\
   6280        &  FTN      &    30    &    $R$    &  $18.86\pm0.05$ & $ 98.0\pm 4.6$ &  1881        &  FTN      &   180    &    $B$    &  $19.62\pm0.03$  & $ 71.9\pm 2.0$\\
   6621        &  FTN      &    60    &    $R$    &  $18.95\pm0.04$ & $ 90.2\pm 3.4$ &  2757        &  FTN      &   120    &    $B$    &  $19.72\pm0.04$  & $ 65.6\pm 2.5$\\
   7110        &  FTN      &   120    &    $R$    &  $19.04\pm0.04$ & $ 83.0\pm 3.1$ &  3511        &  FTN      &   180    &    $B$    &  $19.88\pm0.03$  & $ 56.6\pm 1.6$\\
   7832        &  FTN      &   180    &    $R$    &  $19.10\pm0.03$ & $ 78.5\pm 2.2$ &  5968        &  FTN      &    10    &    $B$    &  $19.94\pm0.18$  & $ 53.6\pm 9.7$\\
   8569        &  FTN      &   120    &    $R$    &  $19.26\pm0.04$ & $ 67.8\pm 2.5$ &  6195        &  FTN      &    30    &    $B$    &  $20.07\pm0.08$  & $ 47.5\pm 3.6$\\
   9258        &  FTN      &   180    &    $R$    &  $19.30\pm0.03$ & $ 65.3\pm 1.8$ &  6510        &  FTN      &    60    &    $B$    &  $20.24\pm0.07$  & $ 40.6\pm 2.7$\\
   9552        &  FTN      &    30    &    $R$    &  $19.25\pm0.07$ & $ 68.4\pm 4.6$ &  6935        &  FTN      &   120    &    $B$    &  $20.28\pm0.05$  & $ 39.2\pm 1.8$\\
   9903        &  FTN      &    60    &    $R$    &  $19.43\pm0.06$ & $ 58.0\pm 3.3$ &  7596        &  FTN      &   180    &    $B$    &  $20.40\pm0.04$  & $ 35.1\pm 1.3$\\
  10414        &  FTN      &   120    &    $R$    &  $19.43\pm0.04$ & $ 58.0\pm 2.2$ &  8401        &  FTN      &   120    &    $B$    &  $20.48\pm0.05$  & $ 32.6\pm 1.5$\\
  11164        &  FTN      &   180    &    $R$    &  $19.50\pm0.04$ & $ 54.3\pm 2.0$ &  9027        &  FTN      &   180    &    $B$    &  $20.55\pm0.05$  & $ 30.5\pm 1.4$\\
  25879        &  AZT33    &   960    &    $R$    &  $20.37\pm0.08$ & $ 24.4\pm1.9$  &  9468        &  FTN      &    30    &    $B$    &  $20.70\pm0.12$  & $ 26.6\pm 3.1$\\
  31242        &  AZT11    &  1800    &    $R$    &  $20.56\pm0.13$ & $ 20.5\pm2.6$  &  9783        &  FTN      &    60    &    $B$    &  $20.82\pm0.09$  & $ 23.8\pm 2.1$\\
  56857        &   LT      &  1860    &    $r'$   &  $21.60\pm0.08$ & $  9.3\pm 0.7$ & 10244        &  FTN      &   120    &    $B$    &  $20.82\pm0.07$  & $ 23.8\pm 1.6$\\
 134128        &  TLS      &  1800    &    $R$    &  $22.1\pm0.3$   & $  5.0\pm1.6$  & 10924        &  FTN      &   180    &    $B$    &  $20.77\pm0.05$  & $ 24.9\pm 1.2$\\
 134490        &  Z1000   &  1620    &    $R$    &  $22.52\pm0.25$ & $  3.4\pm0.9$  &350436        &  Keck     &   785    &    $g'$   &  $24.95\pm0.20$  & $0.449\pm0.091$\\
 306753        &  Z1000   &  4290    &    $R$    &  $> 22.9$       & $< 2.4$        &   310        &  FTN      &    10    &    $V$    &  $20.76\pm0.25$  & $ 20.6\pm 5.3$\\
 350436        &  Keck     &   690    &    $R$    &  $24.17\pm0.14$ & $ 0.74\pm0.10$ &  6025        &  FTN      &    10    &    $V$    &  $19.38\pm0.10$  & $ 73.4\pm 7.1$\\
    382        &  FTN      &    10    &   $i'$    &  $19.62\pm0.16$ & $ 56.1\pm 8.9$ &   260        &  KAIT     &    15    &    $V$    &  $> 17.20$       & $< 547$\\
    651        &  FTN      &    30    &   $i'$    &  $18.05\pm0.03$ & $238.4\pm 6.7$ &   358        &  KAIT     &    45    &    $V$    &  $> 17.70$       & $< 345$\\
   1017        &  FTN      &    60    &   $i'$    &  $17.90\pm0.03$ & $273.7\pm 7.7$ &   532        &  KAIT     &    60    &    $V$    &  $18.60\pm0.70$  & $151\pm136$\\
   1598        &  FTN      &   120    &   $i'$    &  $17.62\pm0.03$ & $354.2\pm 9.9$ &   315        &  KAIT     &    20    &  clear  &  $> 18.00$       & $< 216$\\
   2433        &  FTN      &   180    &   $i'$    &  $17.87\pm0.03$ & $281.3\pm 7.9$ &   468        &  KAIT     &    45    &  clear  &  $18.63\pm0.47$  & $121\pm 66$\\
   3156        &  FTN      &   120    &   $i'$    &  $18.04\pm0.03$ & $240.6\pm 6.7$ &   674        &  KAIT     &    60    &  clear  &  $18.40\pm0.29$  & $150\pm 46$\\
   6100        &  FTN      &    10    &   $i'$    &  $18.41\pm0.07$ & $171.1\pm 11.4$&   958        &  KAIT     &   180    &  clear  &  $18.17\pm0.12$  & $185\pm 22$\\
   6375        &  FTN      &    30    &   $i'$    &  $18.47\pm0.04$ & $161.9\pm 6.1$ & 70327        &  PAIRITEL &  2282    &    $J$    &  $18.70\pm0.19$  & $54.7\pm10.5$\\
   6755        &  FTN      &    60    &   $i'$    &  $18.59\pm0.04$ & $145.0\pm 5.4$ & 81686        &  PAIRITEL &  4363    &    $J$    &  $18.92\pm0.21$  & $44.7\pm9.5$\\
   7325        &  FTN      &   120    &   $i'$    &  $18.61\pm0.04$ & $142.3\pm 5.3$ & 70327        &  PAIRITEL &  2282    &    $H$    &  $17.82\pm0.16$  & $78.0\pm12.4$\\
   8107        &  FTN      &   180    &   $i'$    &  $18.74\pm0.03$ & $126.2\pm 3.5$ & 81686        &  PAIRITEL &  4363    &    $H$    &  $17.81\pm0.15$  & $78.8\pm11.7$\\
   8770        &  FTN      &   120    &   $i'$    &  $18.87\pm0.03$ & $112.0\pm 3.1$ & 70327        &  PAIRITEL &  2282    &    $K$    &  $17.14\pm0.18$  & $94.3\pm17.0$\\
   9650        &  FTN      &    30    &   $i'$    &  $18.99\pm0.06$ & $100.3\pm 5.7$ & 81686        &  PAIRITEL &  4363    &    $K$    &  $16.82\pm0.16$  & $127\pm20$\\
  10035        &  FTN      &    60    &   $i'$    &  $19.02\pm0.04$ & $ 97.5\pm 3.7$ &  6444        &  GeminiN  &    60    &    $r'$   &  $19.10\pm0.03$  & $93\pm3$\\
  10628        &  FTN      &   120    &   $i'$    &  $19.06\pm0.03$ & $ 94.0\pm 2.6$ & & & & & &\\
  11462        &  FTN      &   180    &   $i'$    &  $19.10\pm0.03$ & $ 90.6\pm 2.5$ & & & & & &\\
  60662        &   LT      &  1860    &   $i'$    &  $20.79\pm0.06$ & $ 19.1\pm 1.1$ & & & & & &\\
  77486        &  FTN      &  1591    &   $i'$    &  $20.92\pm0.06$ & $ 17.0\pm 1.0$ & & & & & &\\
\hline
\end{tabular}
 \begin{list}{}{}
  \item[$^{\mathrm{a}}$Midpoint time from the GRB onset time.]
  \item[$^{\mathrm{b}}$Corrected for airmass.]
  \item[$^{\mathrm{c}}$Corrected for Galactic extinction.]
  \end{list}
\end{table*}



\begin{table*}
\centering
  \caption{Equivalent-width measurements for GRB~080603A}
  \label{tab:ew}
  \begin{tabular}{lcccclcccc}
\hline
$\lambda_{\textrm{obs}}$ [\AA] & $\lambda_{\textrm{rest}}$ [\AA] & \hspace*{2mm} $z$ & Feature & EW$_{\textrm{obs}}$ [\AA] &
$\lambda_{\textrm{obs}}$ [\AA] & $\lambda_{\textrm{rest}}$ [\AA] & \hspace*{2mm} $z$ & Feature & EW$_{\textrm{obs}}$ [\AA]\\
\hline
4986.81&1854.72&1.6874&Al~III&$  5.10\pm0.42$ & 7025.48&2612.65&1.6874&FeII*&$  1.80\pm0.06$\\
5008.66&1862.79&1.6874&Al~III&$  4.26\pm0.39$ &        &2614.61&1.6874&FeII*&\\
5325.57&2344.21&1.2714&Fe~II&$  3.72\pm0.15$  & 7039.98&2618.40&1.6874&FeII*&$  0.55\pm0.06$\\
5393.70&2374.46&1.2714&Fe~II&$  2.52\pm0.17$  &        &2621.19&1.6874&FeII*&\\
5413.14&2382.76&1.2714&Fe~II&$  4.46\pm0.15$  & 7051.45&2622.45&1.6874&FeII*&$  0.17\pm0.05$\\
       &2012.17&1.6874&Co~II&                 & 7062.18&2626.45&1.6874&FeII*&$  0.88\pm0.06$\\
5447.13&2026.14&1.6874&Zn~II&$  2.92\pm0.15$  &        &2629.08&1.6874&FeII*&\\
       &2026.27&1.6874&Cr~II&                 & 7075.84&2631.83&1.6874&FeII*&$  1.23\pm0.05$\\
       &2026.48&1.6874&Mg~I&                  &        &2632.11&1.6874&FeII*&\\
5528.99&2056.25&1.6874&Cr~II&$  1.36\pm0.14$  & 7169.21&2796.35&1.5636&MgII&$  1.98\pm0.05$\\
5545.03&2062.23&1.6874&Cr~II&$  2.12\pm0.14$  & 7187.26&2803.53&1.5636&MgII&$  2.82\pm0.07$\\
       &2062.66&1.6874&Zn~II&                 & 7201.86&2683.89&1.6874&VII&$  1.38\pm0.08$\\
5852.00&2175.35&1.6874&Ni~II*&$  0.33\pm0.09$ & 7366.69&3242.93&1.2714&TiII&$  0.59\pm0.06$\\
5875.34&2586.65&1.2714&Fe~II&$  3.67\pm0.11$  &        &2740.36&1.6874&FeII*&\\
5892.87&       &      &    &$ 0.88\pm0.14$   & 7391.43&2747.31&1.6874&FeII*&$  0.65\pm0.07$\\
5906.19&2600.17&1.2714&Fe~II&$  4.77\pm0.11$  &        &2747.85&1.6874&FeII*&\\
6047.64&2249.88&1.6874&Fe~II&$  1.46\pm0.11$  &        &2750.16&1.6874&FeII*&\\
6077.64&2260.78&1.6874&Fe~II&$  1.33\pm0.10$  & 7411.15&2756.56&1.6874&FeII*&$  0.45\pm0.06$\\
6110.07&2382.76&1.5636&Fe~II&$  0.51\pm0.12$  & 7522.96&2796.35&1.6874&MgII&$ 35.39\pm0.07$\\
6274.04&2328.11&1.6874&Fe~II*&$  2.22\pm0.12$ &        &2803.53&1.6874&MgII&\\
       &2333.52&1.6874&Fe~II*& & & & & & \\
       &2338.72&1.6874&Fe~II*& & & & & & \\
6296.24&       &      &     &$ 3.56\pm0.07$ & & & & & \\
6301.42&2344.21&1.6874&Fe~II&$  6.78\pm0.09$ & & & & & \\
       &2345.00&1.6874&Fe~II*& & & & & & \\
6315.59&2349.02&1.6874&Fe~II*&$  1.00\pm0.08$ & & & & & \\
6351.50&2796.35&1.2714&Mg~II&$  7.09\pm0.08$ & & & & & \\
       &2359.83&1.6874&Fe~II*& & & & & & \\
       &2365.55&1.6874&Fe~II*& & & & & & \\
6368.10&2803.53&1.2714&Mg~II&$  7.04\pm0.07$ & & & & & \\
6382.84&2374.46&1.6874&Fe~II&$  5.49\pm0.06$ & & & & & \\
6402.91&2381.49&1.6874&Fe~II*&$ 10.71\pm0.08$ & & & & & \\
       &2382.76&1.6874&Fe~II& & & & & & \\
       &2383.79&1.6874&Fe~II*& & & & & & \\
6423.98&2389.36&1.6874&Fe~II*&$  0.77\pm0.07$ & & & & & \\
6443.15&2396.15&1.6874&Fe~II*&$  1.39\pm0.07$ & & & & & \\
       &2396.36&1.6874&Fe~II*& & & & & & \\
6452.81&2399.97&1.6874&Fe~II*&$  0.63\pm0.07$ & & & & & \\
6482.13&2852.96&1.2714&Mg~I&$  4.82\pm0.12$ & & & & & \\
       &2405.16&1.6874&Fe~II*& & & & & & \\
       &2405.62&1.6874&Fe~II*& & & & & & \\
       &2407.39&1.6874&Fe~II*& & & & & & \\
       &2411.25&1.6874&Fe~II*& & & & & & \\
       &2411.80&1.6874&Fe~II*& & & & & & \\
       &2414.05&1.6874&Fe~II*& & & & & & \\
6631.47&2586.65&1.5636&Fe~II&$  0.21\pm0.06$ & & & & & \\
6666.09&2600.17&1.5636&Fe~II&$  0.55\pm0.06$ & & & & & \\
6886.25&       &      &    &$ 6.66\pm0.12$ & & & & & \\
6926.67&2576.88&1.6874&Mn~II&$  2.84\pm0.07$ & & & & & \\
6945.33&       &      &    &$ 2.11\pm0.05$ & & & & & \\
6953.07&2586.65&1.6874&Fe~II&$  6.98\pm0.05$ & & & & & \\
6981.10&3073.88&1.2714&Ti~II&$  5.99\pm0.06$ & & & & & \\
       &2594.50&1.6874&Mn~II& & & & & & \\
6988.76&2599.15&1.6874&Fe~II*&$  8.97\pm0.05$ & & & & & \\
       &2600.17&1.6874&Fe~II& & & & & & \\
7007.74&2606.46&1.6874&Mn~II&$  2.65\pm0.05$ & & & & & \\
       &2607.87&1.6874&Fe~II*& & & & & & \\
\hline
\end{tabular}
\end{table*}

\end{document}